\documentclass[fleqn,usenatbib]{mnras}

\usepackage{newtxtext,newtxmath}

\usepackage[T1]{fontenc}
\usepackage{ae,aecompl}

\usepackage{graphicx}	
\usepackage{amsmath}	
\usepackage{amssymb}	
\usepackage{comment}
\usepackage{amsfonts}
\usepackage{caption}
\usepackage{pdflscape} 
\usepackage{ulem}
\usepackage{graphicx}
\usepackage{subfig}

\newcommand{\kms}{km~s$^{-1}$}
\newcommand{\unitflux}{erg~cm$^{-2}$~s$^{-1}$}
\newcommand{\unitlum}{erg~s$^{-1}~$}
\newcommand{\Msun}{M$_{\odot}$}
\newcommand{\mbh}{M$_{bh}$}
\newcommand{\ha}{H$\upalpha$}
\newcommand{\hb}{H$\upbeta$}
\newcommand{\hg}{H$\upgamma$}
\newcommand{\hba}{H$\upbeta_{\rm A}$ }
\newcommand{\hbb}{H$\upbeta_{\rm B}$ }
\newcommand{\haa}{H$\upalpha_{\rm A}$ }
\newcommand{\hab}{H$\upalpha_{\rm B}$ }
\newcommand{\nii}{[\ion{N}{ii}] }
\newcommand{\oiii}{[\ion{O}{iii}] }
\newcommand{\chisq}{$\rm \chi^2$}

\title[Gaia16aax]{Extreme variability in an active galactic nucleus: Gaia16aax}

\author[G. Cannizzaro et al.]{
G. Cannizzaro$^{1,2}$ \thanks{E-mail: g.cannizzaro@sron.nl},
M. Fraser$^{3}$,
P. G. Jonker$^{1,2}$,
J. E. Pringle$^{4}$,
S. Mattila$^{5}$,
\newauthor
P. C. Hewett$^{4}$,
T. Wevers$^{4}$,
E. Kankare$^{5}$,
Z. Kostrzewa-Rutkowska$^{1,2,6}$,
{\L}. Wyrzykowski$^{7}$,
\newauthor
F. Onori$^{8}$,
J. Harmanen$^{5}$,
K.E.S. Ford$^{9,10,11}$,
B. McKernan$^{9,10,11}$,
C. J. Nixon$^{12}$\\
$^{1}$SRON, Netherlands Institute for Space Research, Sorbonnelaan, 2, NL-3584CA Utrecht, the Netherlands\\
$^{2}$Department of Astrophysics/IMAPP, Radboud University, P.O. Box 9010, 6500 GL Nijmegen, the Netherlands\\
$^{3}$School of Physics, O'Brien Centre for Science North, University College Dublin, Belfield, Dublin 4, Ireland\\
$^{4}$Institute of Astronomy, University of Cambridge, Madingley Road, Cambridge CB3 0HA, UK\\
$^{5}$Tuorla Observatory, Department of Physics and Astronomy, FI-20014 University of Turku, Finland\\
$^{6}$Leiden Observatory, Leiden University, PO Box 9513, NL-2300 RA Leiden, the Netherlands\\
$^{7}$Warsaw University Astronomical Observatory, Al. Ujazdowskie 4, PL-00-478 Warszawa, Poland\\
$^{8}$Istituto di Astrofisica e Planetologia Spaziali (INAF), Via Fosso del Cavaliere 100, Roma, I-00133, Italy\\
$^{9}$Department of Astrophysics, American Museum of Natural History, Central Park West at 79th Street, New York, NY 10024\\
$^{10}$Department of Science, Borough of Manhattan Community College, City University of New York, New York, NY 10007\\
$^{11}$Physics Program, The Graduate Center, City University of New York, New York, NY 10016\\
$^{12}$Department of Physics and Astronomy, University of Leicester, Leicester LE1 7RH, UK
}

\date{Accepted XXX. Received YYY; in original form ZZZ}

\pubyear{2019}

\begin{document}
\label{firstpage}
\pagerange{\pageref{firstpage}--\pageref{lastpage}}
\maketitle

\begin{abstract}

We present the results of a multi-wavelength follow up campaign for the luminous nuclear transient Gaia16aax, which was first identified in January 2016. The transient is spatially consistent with the nucleus of an active galaxy at z=0.25, hosting a black hole of mass $\rm \sim6\times10^8M_\odot$. The nucleus brightened by more than 1 magnitude in the Gaia G-band over a timescale of less than one year, before fading back to its pre-outburst state over the following three years. The optical spectra of the source show broad Balmer lines similar to the ones present in a pre-outburst spectrum. During the outburst, the \ha~and \hb~emission lines develop a secondary peak. We also report on the discovery of two transients with similar light curve evolution and spectra: Gaia16aka and Gaia16ajq. We consider possible scenarios to explain the observed outbursts. 
We exclude that the transient event could be caused by a microlensing event, variable dust absorption or a tidal encounter between a neutron star and a stellar mass black hole in the accretion disk. We consider variability in the accretion flow in the inner part of the disk, or a tidal disruption event of a star $\geq 1 M_{\odot}$ by a rapidly spinning supermassive black hole as the most plausible scenarios. We note that the similarity between the light curves of the three Gaia transients may be a function of the Gaia alerts selection criteria.
\end{abstract}




\section{Introduction}

Lying at the center of galaxies, supermassive black holes (SMBHs) have an enormous impact on the region surrounding them. Once circumnuclear matter is close enough to the SMBH to fall into the gravitational potential, the black hole will start accreting this matter through the formation of an accretion disk if the material has angular momentum. These Active Galactic Nuclei (AGNs) typically emit over the whole electromagnetic spectrum. Broad (1\,000--20\,000 \kms) and/or narrow (300--1\,000 \kms) highly ionised emission lines are typically present in the optical spectra of AGNs. These lines come from clouds of matter in different regions and at different distances from the central black hole: the Broad Line Region (BLR) and the more distant Narrow Line Region (NLR). The presence or absence of broad and/or narrow lines lead to the classification of AGNs in ``types'': Type 1 when both kinds of lines are present, Type 2 when only narrow lines are present. Intermediate types, depending on the presence of single Balmer lines, exist, e.g. type 1.9 characterised by the presence of a broad \ha~emission line but with an absent broad \hb~\citep{osterbrock81}. 
The unified model explains this dichotomy in terms of our viewing angle to the AGN, where a parsec-scale dusty torus may partially or totally obscure our view of the central engine and the BLR \citep{antonucci93,urrypadovani95}.

AGN are intrinsically variable objects, with stochastic variations in brightness of order of 20\% over time-scale of months to years and larger variations happening on longer timescales \citep{macleod10,macleod12}. In recent years, more extreme examples of Quasar\footnote{We use the term Quasar or QSO to mean the most luminous fraction of the AGN population, with bolometric luminosity $\rm L\gtrsim10^{44}$\unitlum} variability have been discovered, e.g.  \citet{graham16} and \citet{rumbaugh18} found large  samples of objects that showed variability of more than 1 magnitude over a timescale of several years in an archival search in the Catalina Real-Time Transient Survey, Sloan Digital Sky Survey and Dark Energy Survey. \citet{lawrence16} investigated the nature of a sample large amplitude nuclear transients discovered during the Pan-STARSS $\rm 3\pi$ survey. The majority of these objects are classified as hypervariable AGNs. In some cases, AGNs will show slow and significant optical variability of more than 1 magnitude over several years, accompanied by spectral variability (e.g \citealt{matt03,lamassa15,macleod16}). These so-called "changing look AGNs" have been observed transitioning between different AGN types, with the (dis)appearance of broad emission lines (typically the Balmer series). While various physical mechanisms have been proposed to explain this extreme variability, such as variable obscuration across the line of sight, microlensing or accretion disk instabilities, no single explanation has been found. \citet{merloni15} proposes as an explanation the occurrence of a Tidal Disruption Event (TDE) in the nucleus of the galaxy. A TDE happens when a star passes so close to a SMBH that the tidal forces of the SMBH are stronger than the star's self-gravity. This leads to the (partial) disruption of the star \citep{hills75, rees88, evans89}. The disruption of the star and the subsequent accretion of roughly half of the stellar material (the other half will be expelled on unbound orbits) gives rise to a short and luminous flare that typically peaks in the UV or soft X-rays. The bolometric light curve decay of the transient event is expected to follow the fallback rate of the material with a powerlaw decline $\rm \propto t^{-5/3}$ on a timescale of some months up to a year \citep{evans89, cannizzo90,lodato09}. TDEs have been invoked as a possible explanation also for numerous objects in the large samples described in \citet{graham16,lawrence16,rumbaugh18}.

Here, we report the follow up observations of Gaia16aax, a transient event happening at the center of a galaxy hosting a known QSO (SDSS J143418.47+491236.5, \citealt{sdss14}) at z=0.25, discovered by the Gaia Science Alerts project (GSA, \citealt{hodgkin13}). The transient was given the designation AT2016dbt\footnote{https://wis-tns.weizmann.ac.il/object/2016dbt} in the Transient Name Server (TNS). The center of the galaxy brightened by $\sim$1 mag (in the optical) over the course of 1 year, before fading back to its pre-outburst state over more than two years. Variability of similar amplitude is detected in X-rays. Together with the photometric variability, the object undergoes a dramatic change also in its spectroscopic characteristics. The broad Balmer lines (\ha~ and \hb), already present in the pre-outburst spectrum, during the outburst show a significantly different morphology, with two peaks of different intensity and separated by $\sim$100 \AA. The initial classification spectrum showed a strong blue continuum with broad, double-peaked Balmer lines. After the initial classification as an AGN outburst, we started a multiwavelength follow-up campaign with optical and NIR photometry, optical spectroscopy, and X-ray observations.

In Section \ref{sec:obs} we describe both the available, pre-outburst data and our own follow-up data, in Section \ref{sec:analysis} we describe our analysis of the photometric data, the Spectral Energy Distribution (SED) and the fit to the emission lines, finally in Section \ref{sec:discussion} we discuss our results in the broader framework of extreme AGN variability and we discuss some possible scenarios to explain the outburst.

Throughout the paper we assume a cosmology with $H_{\rm 0}$=67.7 km s$^{-1}$ Mpc$^{-1}$, $\Omega_{\rm M}$=0.309, $\Omega_{\rm \Lambda}$=0.691, consistent with \citet{planck15}.


\section{Observations}
\label{sec:obs}
\subsection{Gaia data and discovery}
\label{sec:gaia_disc}
\begin{figure}
    \includegraphics[width=1\columnwidth]{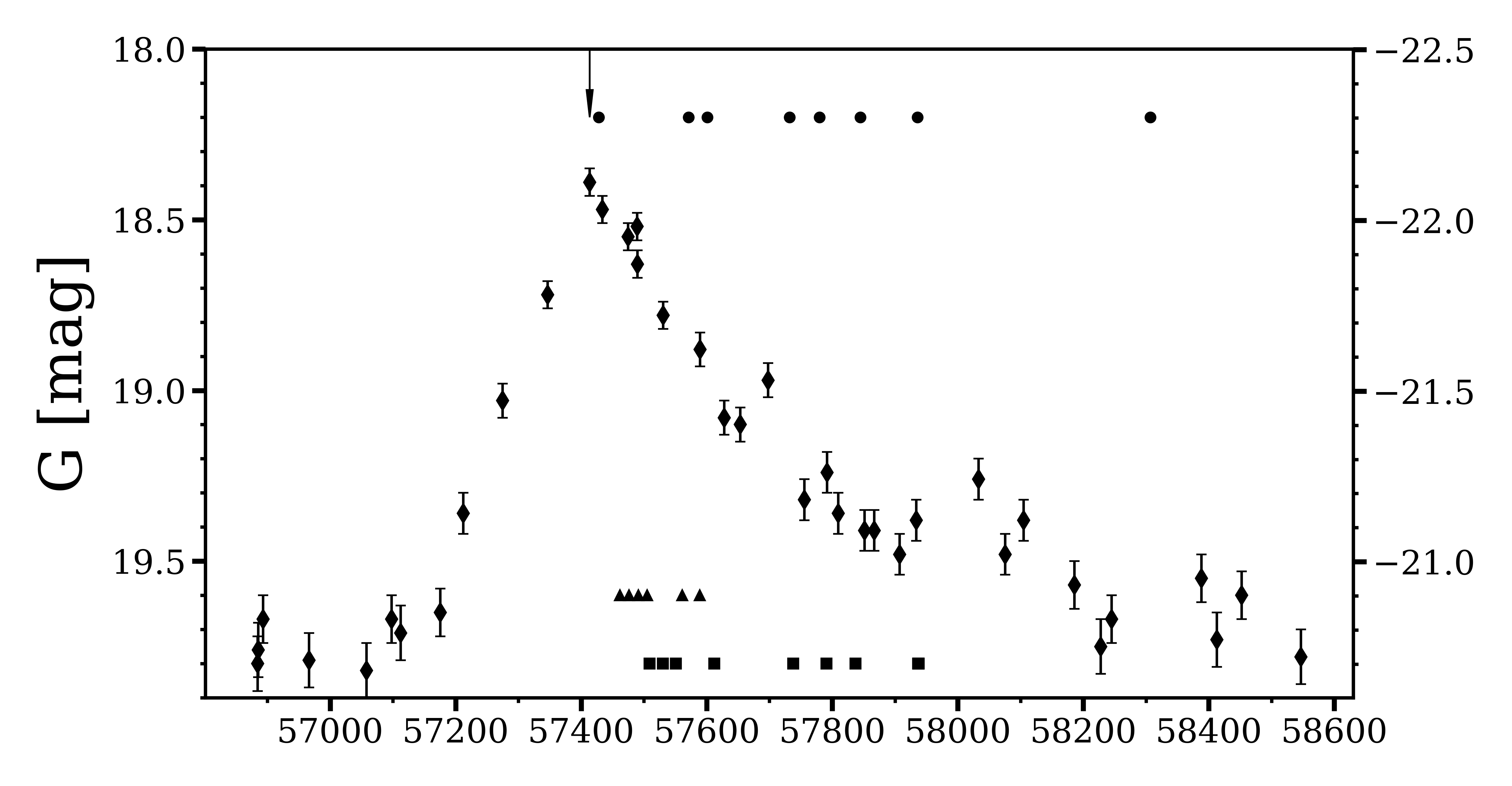}
    \caption{Light curve from the {\it Gaia} satellite for Gaia16aax. The arrow indicates the date the transient was alerted on. Magnitudes are in the {\it Gaia} {\it G} broad band filter \citep{jordi10}. The black circles at the top indicate the epochs at which an optical spectrum was taken. The black triangles and squares at the bottom show the epochs at which optical and NIR images were taken, respectively. The two y-axes are apparent (left axis) and absolute (right axis) magnitudes.}
    \label{fig:gaia_lc}
\end{figure}

Gaia16aax was first alerted on by the Gaia Photometric Science Alerts program\footnote{http://gsaweb.ast.cam.ac.uk/alerts/alert/Gaia16aax} on 2016 January 26. The transient is $\sim0.02\arcsec$ from the position of the host galaxy in the Gaia Data Release 2 (DR2, \citealt{gaiadr2}) and $\sim0.1\arcsec$ from the position of the host galaxy in the Sloan Digital Sky Survey (SDSS, \citealt{sdss14}). As shown in \citet{kostrzewa18}, where nuclear transients detected by Gaia were matched astrometrically with SDSS sources, the astrometry provided by the Gaia Photometric Science Alerts system is  accurate to $\sim0.1\arcsec$. We hence regard Gaia16aax as consistent with having no offset from the host nucleus. The {\it Gaia} lightcurve is shown in Fig.\ref{fig:gaia_lc}: it shows a steady, smooth rise up to its maximum in less then a year and a slower decay on a timescale of more than two years to its pre-outburst, quiescent level. The apparent magnitude at peak in the Gaia broad G filter (which corresponds to white light, see \citealt{jordi10}) is $\rm G=18.39$, which corresponds to an absolute magnitude $\rm M_G=-22.11$, given a source redshift z$\simeq$0.248 and a luminosity distance $\rm D_L=1.26$ Gpc, assuming no extinction. The date of the alert issue corresponds with the date at which, according to the Gaia lightcurve, the flare reached its peak emission.

\subsection{Sloan Digital Sky Survey and Pan-STARRS}

\begin{figure}
    \includegraphics[width=1\columnwidth]{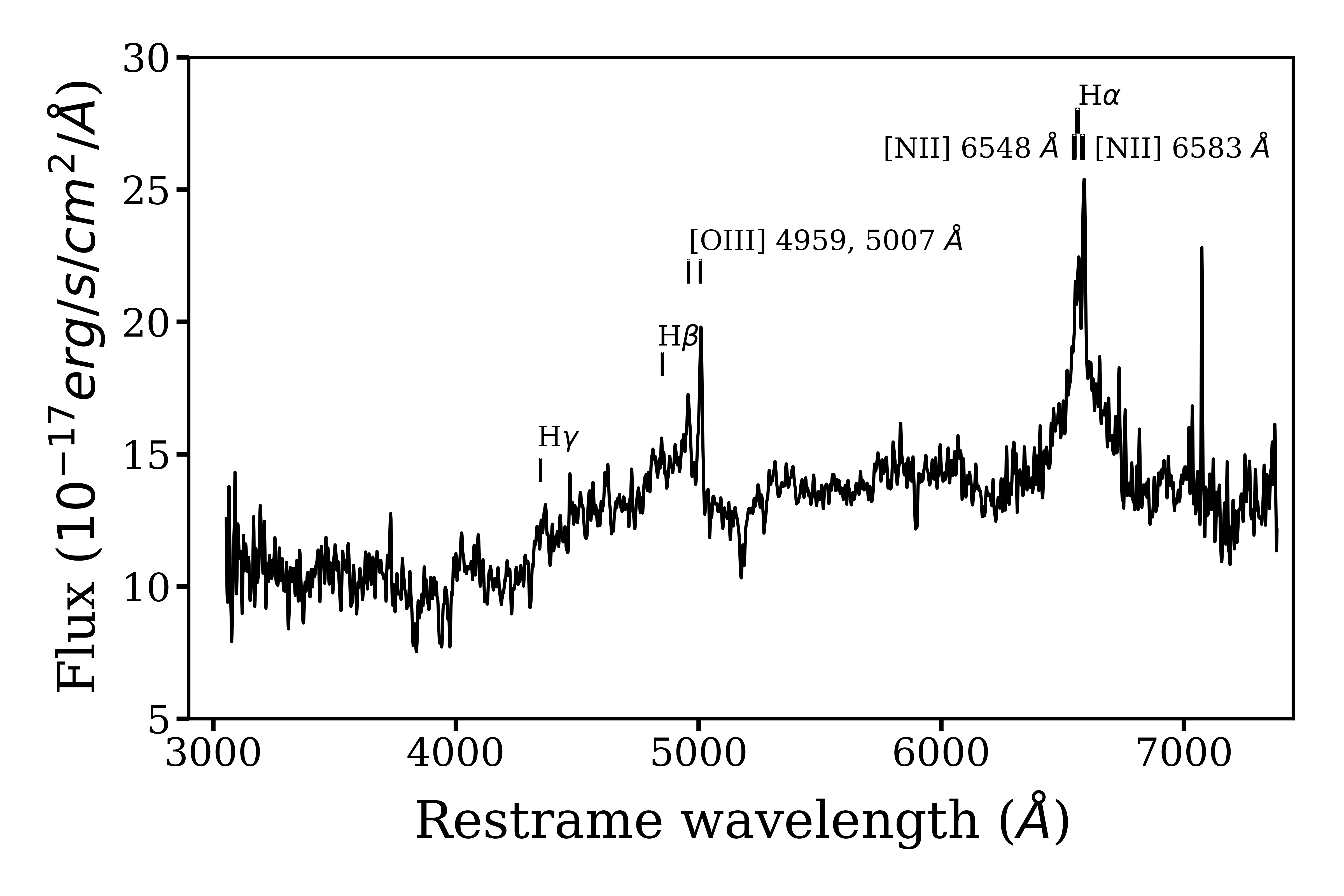}
    \caption{Archival spectrum from the SDSS legacy survey. The object shows broad Balmer emission lines and narrow forbidden oxygen and nitrogen lines.}
    \label{fig:sdss_spec}
\end{figure}

The source is present in SDSS Data Release 14 as SDSS J143418.47+491236.5. It was imaged on 2002 May 8\footnote{from here onwards all times are in UTC.} and morphologically classified as a galaxy with r=17.76$\pm$0.01 mag. The source was then observed spectroscopically during the SDSS legacy survey on 5 July 2002 and classified as a broadline quasar at redshift z=0.24853$\pm$0.00005 \citealt{sdss14}. The SDSS spectrum is shown in Fig.\ref{fig:sdss_spec}. The Point Spread Function (PSF)-fitting magnitudes are reported in Table \ref{tab:photometry}. The object is also present in Pan-STARSS Data Release 2 as PSO J143418.490+491236.529 (Object ID 167052185770922854). The PSF magnitudes for the filters (\textit{g,r,i,z}) are similar to, though not formally consistent with the ones reported in SDSS. The different values are likely due to the differences in the PSF modeling used in the two surveys.

\subsection{Catalina Sky Survey}

\begin{figure}
    \includegraphics[width=1\columnwidth]{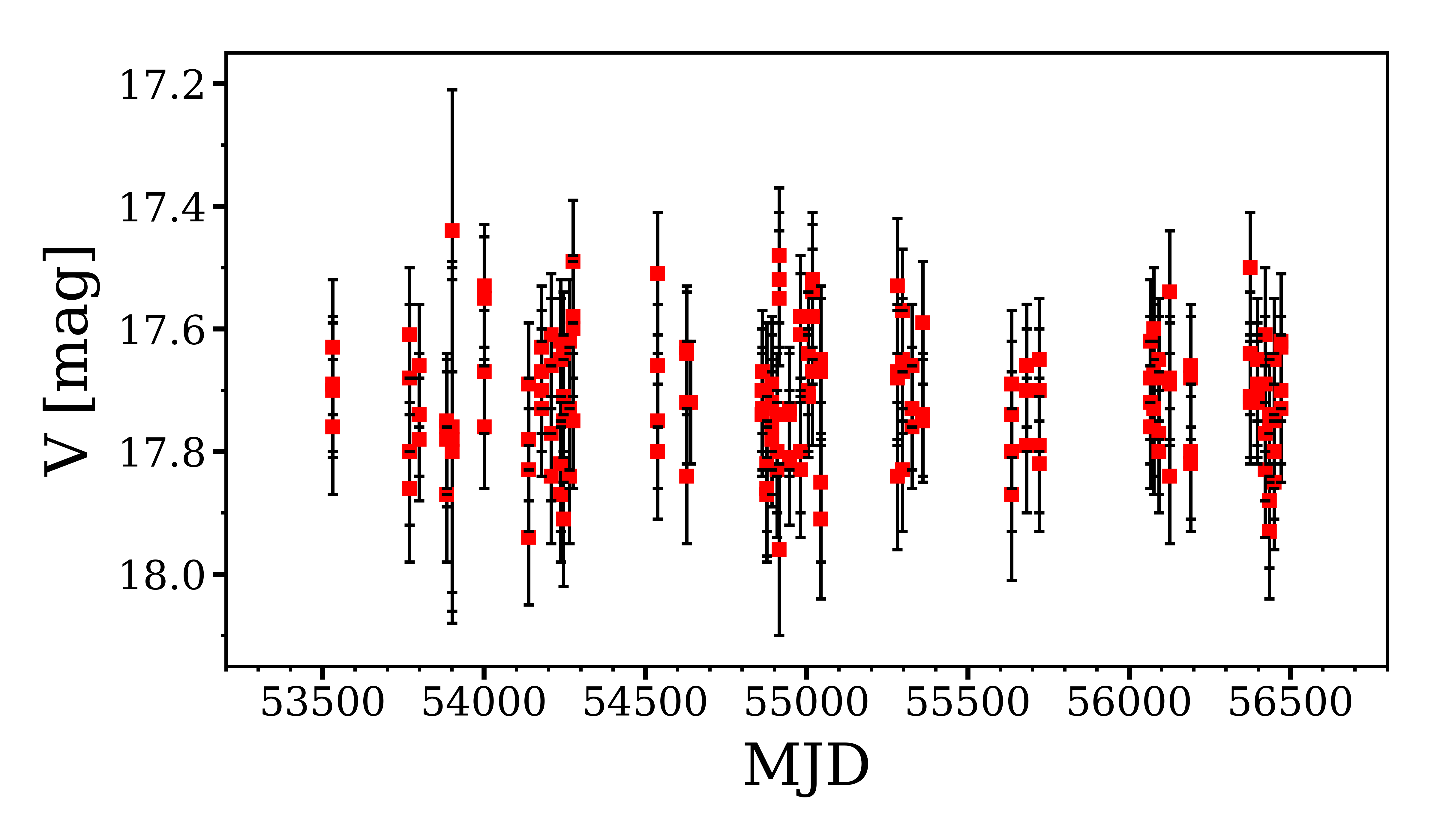}
    \caption{Historical lightcurve from the Catalina Sky Survey. In the period covered by the survey the source has not exhibited an outburst similar to Gaia16aax.}
    \label{fig:cat_lc}
\end{figure}

The source has been observed by the Catalina Sky Survey (CSS)\footnote{The CSS survey is funded by the National Aeronautics and Space Administration under Grant No. NNG05GF22G issued through the Science Mission Directorate Near-Earth Objects Observations Program.}\citep{Drake09} over a period of eight years (from June 2005 to June 2013, source ID CSS J143418.6+491236). As shown in Fig. \ref{fig:cat_lc}, the object has not shown an enhanced emission state like Gaia16aax during the eight years of CSS monitoring.

\subsection{2MASS}
The source is present in the Two Micron All-Sky Survey (2MASS) as 2MASS J14341848+4912366. It has been observed once on 2000 May 07 in the three Near Infra-Red bands {\it J}, {\it H} and {\it K$_s$}, with magnitude values $\rm J=15.96\pm0.09$, $\rm H=15.47\pm0.14$ and $\rm K_s=14.64\pm0.10$. 

\subsection{(NEO)WISE}
The source has been observed in the infrared band by both the WISE and the NEO-WISE (Near-Earth Objects Wide-field Infrared Survey Explorer, \citealt{mainzer11}) phases of the {\it WISE} satellite between January 2014 and June 2017. This period has overlap with both the Gaia data (see Sec.\ref{sec:gaia_disc}) and our own follow-up data. The WISE satellite provides observations in four filters W1, W2, W3 and W4 (3.4, 4.6, 11.6 and 22.1 $\upmu$m respectively), while the reactivation mission NEO-WISE only provides values for the W1 and W2 filters, in which it takes 10-20 single exposures for each epoch of observation. We selected only data points with photometric flags A or B (for which the minimum flux signal-to-noise ratio is above 10 and 3, respectively). The NEO-WISE magnitudes in the W1 and W2 bands in the quiescent phase (before MJD 57200) are compatible with the ones measured during the original WISE mission \citep{wright10}: W1 = 13.783$\pm$0.025 and W2 = 13.17$\pm$0.02 observed on 21 Jul 2010. The WISE magnitudes in the other two filters (not available during the NEO-WISE phase of the mission) are W3 = 10.93$\pm$0.08 and W4 = 8.72$\pm$0.32.

\subsection{UV and X-ray observations}

The source was observed with all the instruments on board the Neils Gehrels Swift Observatory (\citealt{gehrles04}, \textit{Swift} from here on) on 9 and 14 September 2006. UV data were analysed with the \textit{Swift} task \texttt{UVOTSOURCE}, using a $5''$ aperture to estimate the source brightness and a $50''$ aperture, centered on a nearby empty region of sky, to estimate the background levels. Combining the two epochs of observations we derived that the source has UV magnitudes UVW1=21.5$\pm$0.2, UVM2=21.87$\pm$0.15, and UVW2=21.7$\pm$0.1 (in the AB magnitude system).

In X-rays the source was observed for a total exposure time of 14.2 ks. X-Ray Telescope (XRT) data was analysed using the online XRT product builder \citep{evans09}. The observed flux in the 0.3 -- 10.0 keV range is $\rm 2.5\pm0.9\times10^{-13}$\unitflux, obtained modeling the data with a power-law with photon index $\upalpha\sim1.4$. This, at the luminosity distance reported previously, corresponds to a luminosity $\rm L_X=3.7\pm0.9\times10^{42}$\unitlum.

We observed the source again with \textit{Swift} on 18 July 2019. The UV and X-ray data was reduced and analysed in the same way as the archival data. We derived UV magnitudes UVW1 = $19.45\pm0.22$, UVM2=$19.38\pm0.24$ and UVW2=$19.52\pm0.23$. The 0.3 -- 10.0 keV X-ray luminosity is $\rm L_X=(1.7\pm0.7)\times10^{43}$\unitlum. The object is therefore still bright in the UV and X-rays, with respect to the archival values.

We observed Gaia16aax also with XMM-Newton, for which we were awarded a Director's Discretionary Time observation. The observation started on June 30, 2016. During the observations, the pn camera was operated in Full Window mode, providing a time resolution of 73.4 ms. Both the MOS detectors were operated in Partial Window mode, where the Small Window option was used, implying that only the central 100$\times$100 pixels of 1.1\arcsec\, each were read out. This yields a time resolution of 0.3 seconds. The source flux is too low to yield meaningful Reflection Grating Spectrometer data.

We run the default {\sc SAS} v17 (20180620) tool {\sc xmmextractor} under the HEASOFT {\sl ftools} software version 6.24 to extract source light curves and spectra. After filtering for periods of high background, we are left with a total exposure for the pn and MOS1 and MOS2 of 46.1, 59.7 and 59.4 ksec, respectively. We use the {\sc SAS} command {\sc epatplot} to assess if photon pile--up is important during our observation. We conclude pile--up is not important, as expected given the overall number of photons detected and given that the light curve does not show large flares.

The pn source spectrum was extracted from a circular region with radius of 30\arcsec\, centered on the known optical source position. Background photons were extracted from a source-free circle with a radius of 59.59\arcsec\, centered on Right Ascension 218.527 and Declination 49.2065. The MOS source spectra were extracted from circular regions with radius of 30\arcsec\, centered on the known optical source position. The MOS background spectra are obtained from an annulus centered on the source position, with inner and outer radius of 330 and 660\arcsec, respectively. The extracted spectra are rebinned to yield a minimum of 30 counts per bin.

We fitted the spectra with a powerlaw modified by Galactic foreground extinction. The reduced $\chi^2$ of the fit is 1.24 for 250 degrees of freedom. The normalisation for the pn and MOS power laws are within a few percent of each other, and we tie the other parameters (N$_H$ and the powerlaw index) to require them to be the same. The best--fitting power law index is 1.85$\pm$0.01, while the foreground extinction is 2$\times 10^{20}$ cm$^{-2}$. The 0.3--10 keV absorbed and unabsorbed flux is $(1.4\pm0.1)\times 10^{-12}$, $(1.5\pm0.1)\times 10^{-12}$\unitflux, respectively, corresponding to a luminosity $(2.8\pm0.1)\times 10^{44}$\unitlum.

Simultaneously to the X-ray observation, we obtained photometry in the UV filters {\it UVM2} and {\it UVW1} and optical filter Johnson {\it U} with the Optical Monitor (OM) on board of \textit{XMM-Newton}. The resulting magnitudes are $\rm UVM2=17.93\pm0.14$, $\rm UVW1=18.15\pm0.04$ and $\rm U=18.32\pm0.03$, in the AB magnitude system.

\subsection{NOT spectroscopic observations}
\label{sec:Spec}
\begin{figure*}
    \includegraphics[scale=0.5, angle = 0]{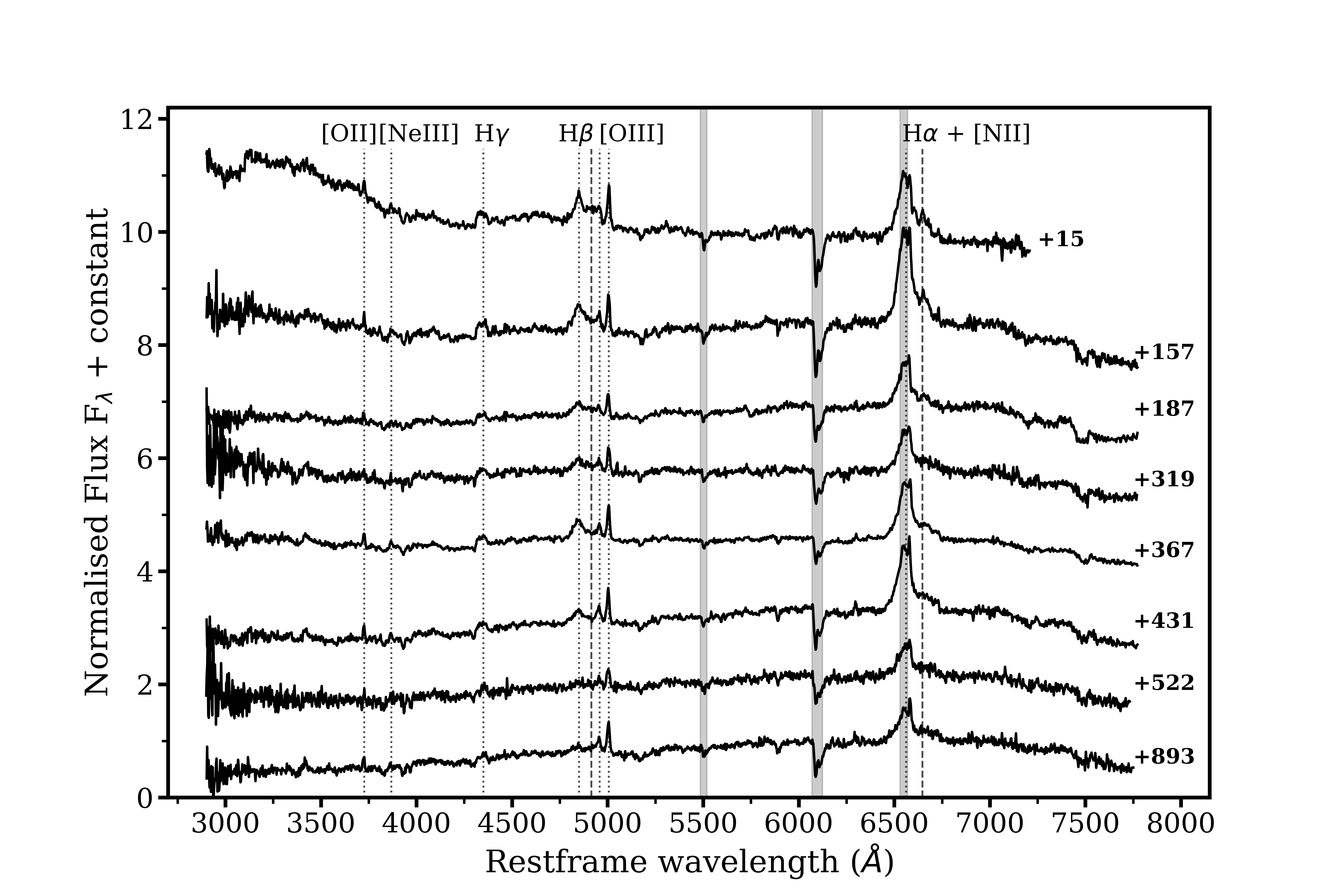} 
    \caption{Eight optical spectra taken with ALFOSC at the NOT, shifted to the rest-frame of the host galaxy. For each spectrum we annotate to the right the number of days passed since the publishing of the Gaia Science Alert (2016 January 26, see Sec.\ref{sec:gaia_disc}). Note the large gap in time between the first and second spectroscopic observation. The dotted lines indicate the wavelengths of the main emission lines, the dashed lines indicate the secondary components (red wings) of the \ha~and \hb~lines. The grey bands represent wavelengths affected by the telluric absorption, one of which affects the \ha~line region.}
    \label{fig:gaia_spec}
\end{figure*}

We started a follow-up campaign of Gaia16aax using the Alhambra Faint Object Spectrograph (ALFOSC) using grism \#4 (3200 - 9600~\AA, (R$\sim$360) for a slit of 1.0\arcsec) at the Nordic Optical Telescope (NOT), in the framework of the NOT Unbiased Transient Survey\footnote{\texttt{http://csp2.lco.cl/not/}}. For all observations the slit was placed at parallactic angle. We obtained 7 epochs of optical spectroscopy between 2016 February 09 and 2017 July 01 (see Table \ref{tab:obs_spec}). All observations were reduced using \texttt{foscgui}\footnote{Foscgui is a graphic user interface aimed at extracting SN spectroscopy and photometry obtained with FOSC-like instruments. It was developed by E. Cappellaro. A package description can be found at http://sngroup.oapd.inaf.it/foscgui.html} which is a pipeline based on standard \texttt{iraf} data reductions procedures: flat field and bias correction, cosmic-ray cleaning, wavelength and flux calibration with arc lamps and standard stars and telluric line correction \citep{tody86}. The pipeline also performs second order contamination removal for the spectra following \citet{stanishev07}. The sequence of ALFOSC spectra is shown in Fig.\ref{fig:gaia_spec}.

\begin{figure}
    \includegraphics[width=1\columnwidth]{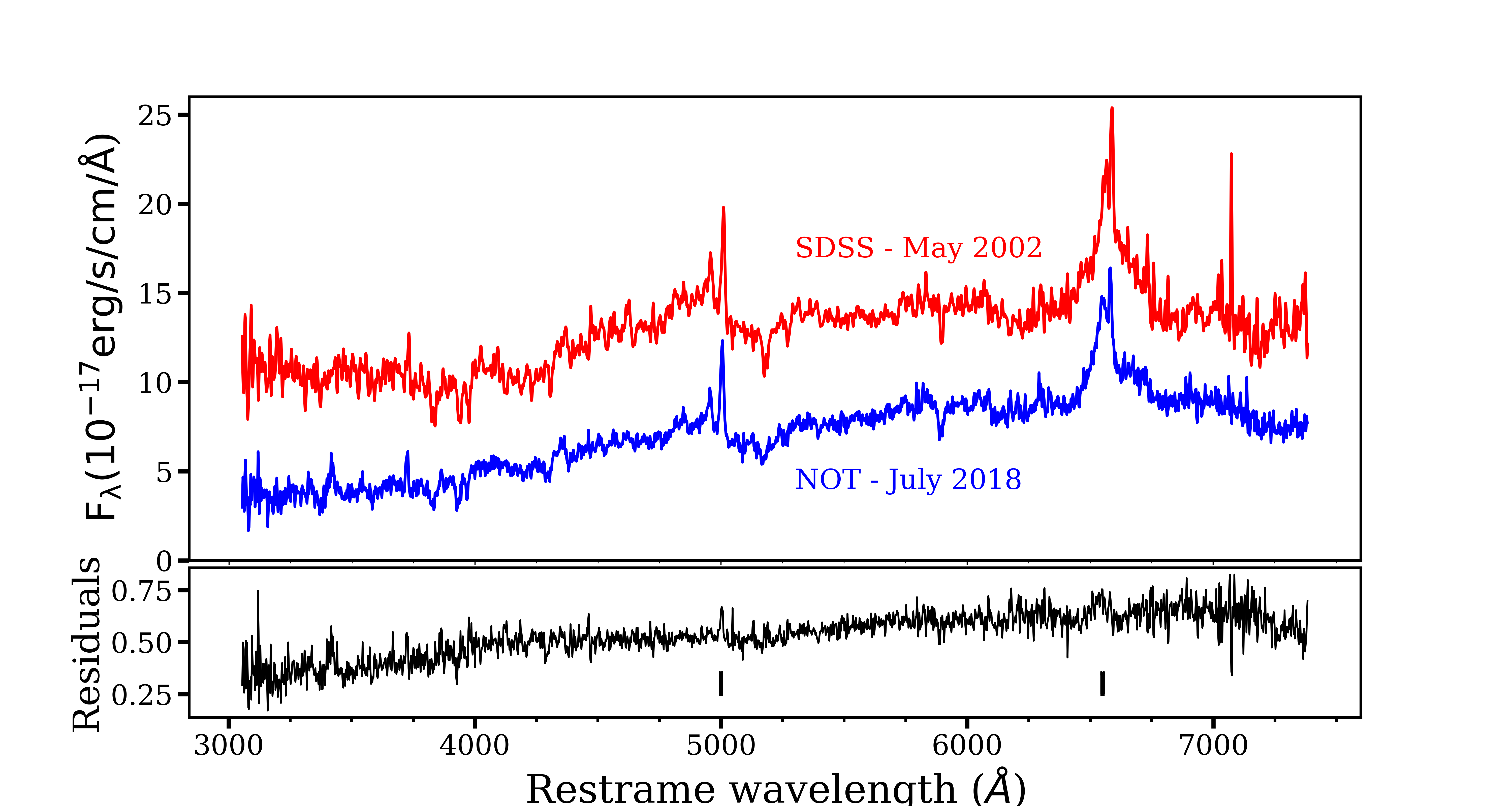}
    \caption{Comparison between the SDSS pre-outburst spectrum (in red) and the last spectrum of our follow-up (July 2018, in blue). The division of the two spectra results in a nearly featureless spectrum, shown in the bottom panel. In the residual spectrum there is still a trace of the \ha~and \oiii  emission lines (marked in the bottom panel).}
    \label{fig:spec_div}
\end{figure}

On 2018 July 08 we took one last spectrum to verify that the object had returned to quiescence as suggested by its Gaia lightcurve (see Fig.\ref{fig:gaia_lc}). Indeed, this spectrum appears to be similar to the pre-outburst SDSS spectrum and dividing the two spectra results in an almost featureless spectrum but with a slight slope. The slope and the presence of residual \ha~and \oiii~lines can both be introduced due to the use of different spectrographs by SDSS (optical fibers) and NOT (long-slit): an optical fiber will collect more light from the whole galaxy, while the slit is centered on the nuclear region, hence the SDSS spectrum will have an excess of light coming from the circumnuclear region (see Fig. \ref{fig:spec_div}).

\begin{table}
	\centering
 	\caption{ALFOSC/NOT spectroscopic observations}
 	\label{tab:obs_spec}
 	\begin{center}
	\begin{tabular}{lcccl}
		\hline
		MJD$^{(1)}$ & UTC Date  & exposure time   & slit   &airmass  \\
        	{[d]}		 &          & [s]             &[$''$]     \\
        \hline
		57\,428.27 & 2016 Feb 10 & 2400       & 1.0 &	1.07 \\ 
		57\,571.06 & 2016 Jul 02 & 1800       & 1.0 &	1.56 \\
		57\,600.97 & 2016 Jul 31 & 2700       & 1.0 &	1.49 \\
		57\,732.25 & 2016 Dec 10 & 1800       & 1.0 &	1.76 \\
		57\,780.26 & 2017 Jan 27 & 2400       & 1.0 & 1.11 \\
		57\,845.15 & 2017 Apr 02 & 2700       & 1.0 &	1.07 \\
		57\,935.94 & 2017 Jul 01 & 2000       & 1.3 &	1.10 \\
		58\,307.01 & 2018 Jul 08 & 2700       & 1.0 & 1.32 \\
		\hline
\end{tabular}
\end{center}
\textit{Note.}(1) Modified Julian Day of observations.
\end{table} 
\subsection{Photometric observations}
Besides our spectroscopic follow-up, we started a series of photometric observations to monitor the decay of the transient to complement the lightcurve from Gaia. These photometric observations were also performed in the framework of the NOT Unbiased Transient Survey using NOTCam (Nordic Optical Telescope near-infrared Camera and spectrograph) at the NOT for the NIR bands J, H and K$_s$. In addition we employed the Optical Wide Field camera (IO:O) at the Liverpool Telescope (LT) using the Sloan \textit{ugriz} and Bessel B and V filters. After the source went back to its pre-outburst emission level, we obtained a final set of observations in all necessary filters to subtract the host-galaxy/pre-outburst contribution to the flux detected in outburst. NOTCAM images were reduced using a modified version of the external NOTCAM \texttt{iraf} package\footnote{http://www.not.iac.es/instruments/notcam/guide/observe.html\#reductions} (version 2.5) and their zero points calibrated using 2MASS (Two Micron All-Sky Survey, \citealt{Skrutskie06}) catalogue. Optical images were reduced through the Liverpool Telescope pipelines and their zero points were calculated using stars in SDSS. For the calculation of the zero points in Bessel B and V, we applied the filter transformation equations found in \citet{jester05}.

For each epoch of observation we performed differential photometry to measure the magnitude of the transient. For this we subtracted a template image, taken when the object was back at its pre-outburst emission level, from our scientific images, using \texttt{HOTPANTS}\footnote{https://github.com/acbecker/hotpants}~\citep{hotpants}. This program uses the algorithm from \citet{alard98} for the subtraction of images taken under different seeing conditions: it applies a spatially varying kernel to one of the two images to match the PSF of the other one and then subtracts the template image to obtain a final image in which only the transient is visible, while all constant luminosity sources will have been subtracted.

On these subtracted images we performed aperture photometry using the \texttt{iraf} task \texttt{APPHOT} with variable apertures depending on the seeing conditions.
Uncertainties on the magnitudes are obtained by adding in quadrature the photometric error from the aperture photometry and the standard deviation due to the scatter in the sources used to calculate the zero points. We did not apply band-pass corrections as our uncertainties are dominated by systematic errors due to the image-subtraction process.

\subsection{Observations of other Gaia transients}
\label{sec:gaia_objects}

In the first months of 2016, other two objects with very similar characteristics to Gaia16aax were discovered by the Gaia Photometric Science Alerts pipeline: Gaia16ajq\footnote{https://gsaweb.ast.cam.ac.uk/alerts/alert/Gaia16ajq/} (AT2016dvz on the TNS) and Gaia16aka\footnote{https://gsaweb.ast.cam.ac.uk/alerts/alert/Gaia16aka/} (AT2016dwk on the TNS).\\
Gaia16ajq was discovered on 2016 March 31, as an increase in brightness by $\sim$1 mag in the nucleus of an active galaxy at z$\sim$0.28. The galaxy is present in SDSS and it is classified as a QSO/starburst galaxy. 
Gaia16aka was discovered on 2016 April 4 also as an increase in brightness of $\sim$1 mag of the nucleus of a galaxy at z$\sim$0.31. This galaxy is present in SDSS but an archival spectrum is not available, therefore we were unable to determine if the galaxy hosts an AGN. 

\begin{figure}
    \includegraphics[width=1\columnwidth]{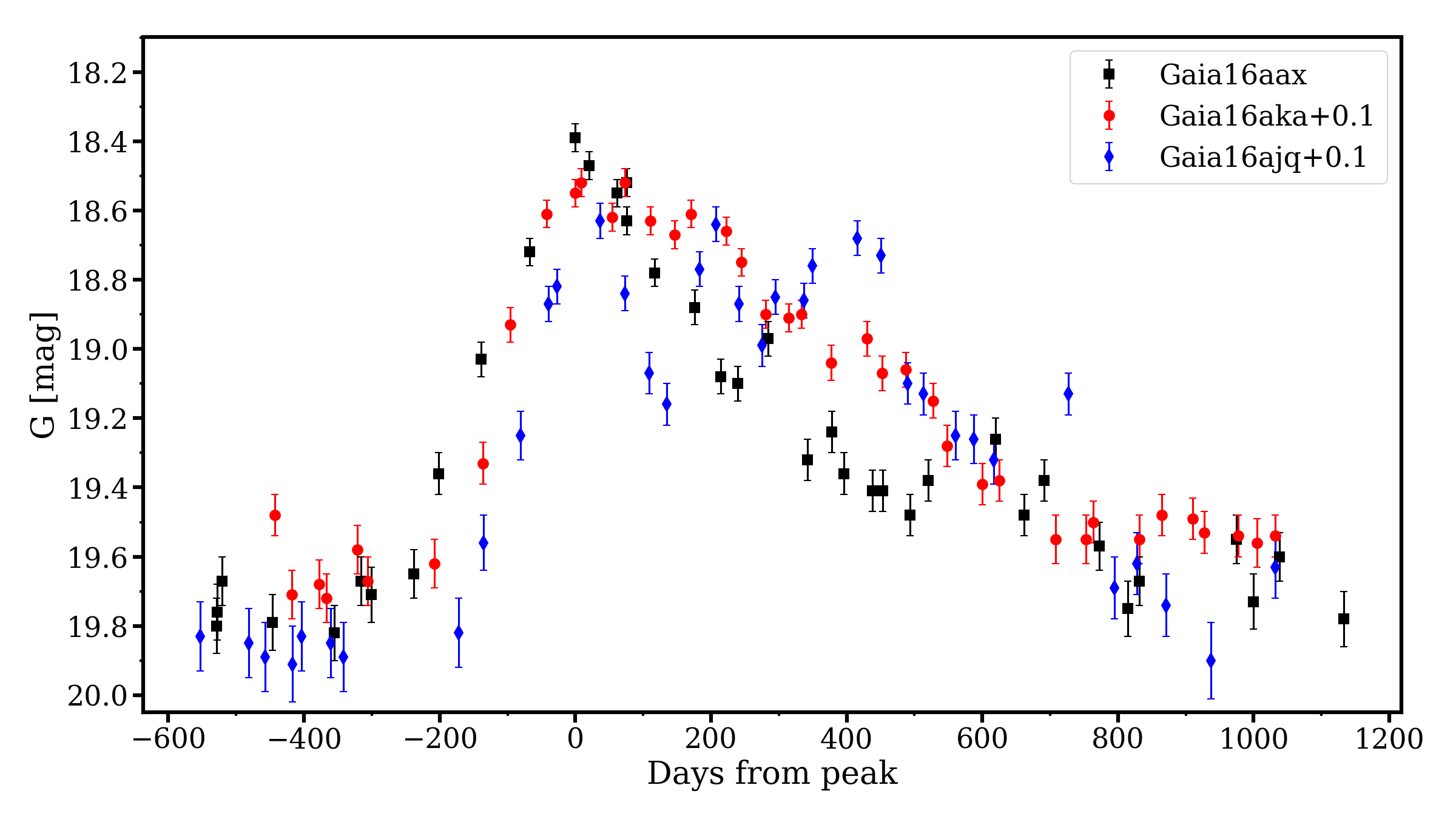}
    \caption{Light curves of the three similar sources found by the Gaia Alerts system: Gaia16aax (black squares), Gaia16ajq (blue diamonds) and Gaia16aka (red circles). For the sake of comparison, all three lightcurves have been shifted vertically and horizontally. The 0 on the x-axis is the date at which each object reaches its peak emission.}
    \label{fig:gaia16_lc}
\end{figure}

In Fig.\ref{fig:gaia16_lc} the Gaia lightcurves of Gaia16ajq and Gaia16aka, together with the one of Gaia16aax, are plotted. The lightcurves have been aligned using the time of peak as point of reference and shifted vertically by 0.1 mag for comparison. All three objects show a similarly shaped rise to peak emission on the same timescale of a few hundreds of days. The lightcurve decay of the three objects is less similar than the rise time. All three objects show bumps in the decay part of the lightcurve, with Gaia16ajq showing two very prominent ones at $\sim$200 and $\sim$400 days after peak.

\begin{figure}
    \includegraphics[width=1\columnwidth]{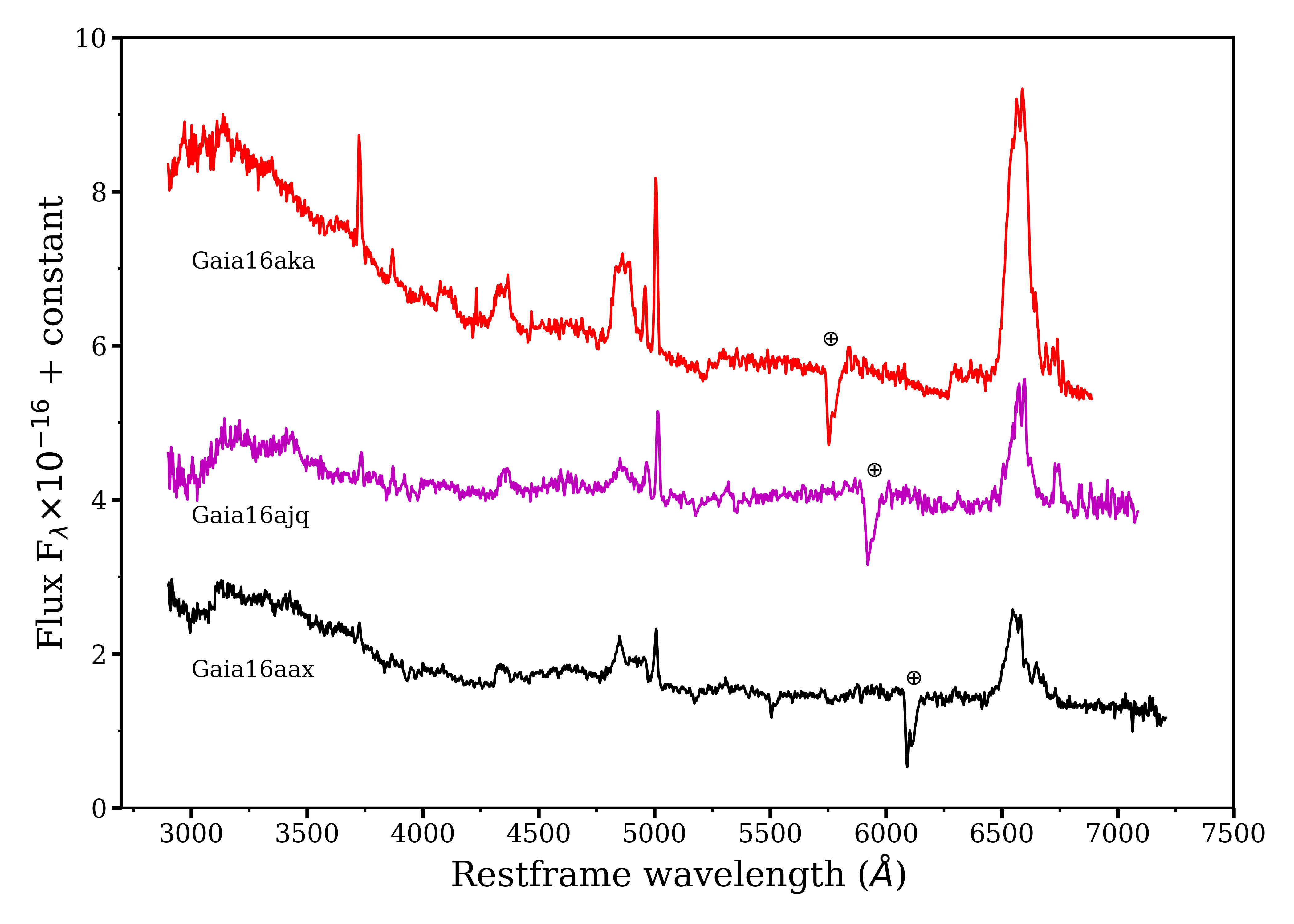}
    \caption{Spectra of the three similar Gaia objects: Gaia16aka (at the top, in red), Gaia16ajq (in the middle, in magenta) and Gaia16aax (at the bottom, in black). With $\oplus$ we indicate telluric absorption bands.}
    \label{fig:gaia16_spec}
\end{figure}

The three objects show similarities also in their spectra. For Gaia16ajq and Gaia16aka we only have a classification spectrum, taken with the NOT a few days after discovery. Both spectra and the first spectrum of Gaia16aax, for comparison, are shown in Fig. \ref{fig:gaia16_spec}. All three objects show a blue bump - with Gaia16aka showing the most prominent one - broad Balmer lines and narrow forbidden lines typical of AGNs. In all three spectra the continuum between \hg~and \hb~is high, this could be due to the presence of unresolved lines due to the Bowen fluorescence mechanism or to a forest of unresolved Fe lines.

While the \ha~and \hb~lines in Gaia16aax showed a very distinct double-peaked shape during the outburst, both Gaia16ajq and Gaia16aka do not show the same morphology in their spectra.



\section{Analysis and results}
\label{sec:analysis}


\subsection{Photometric and SED analysis}

\begin{figure}
    \includegraphics[width=1\columnwidth]{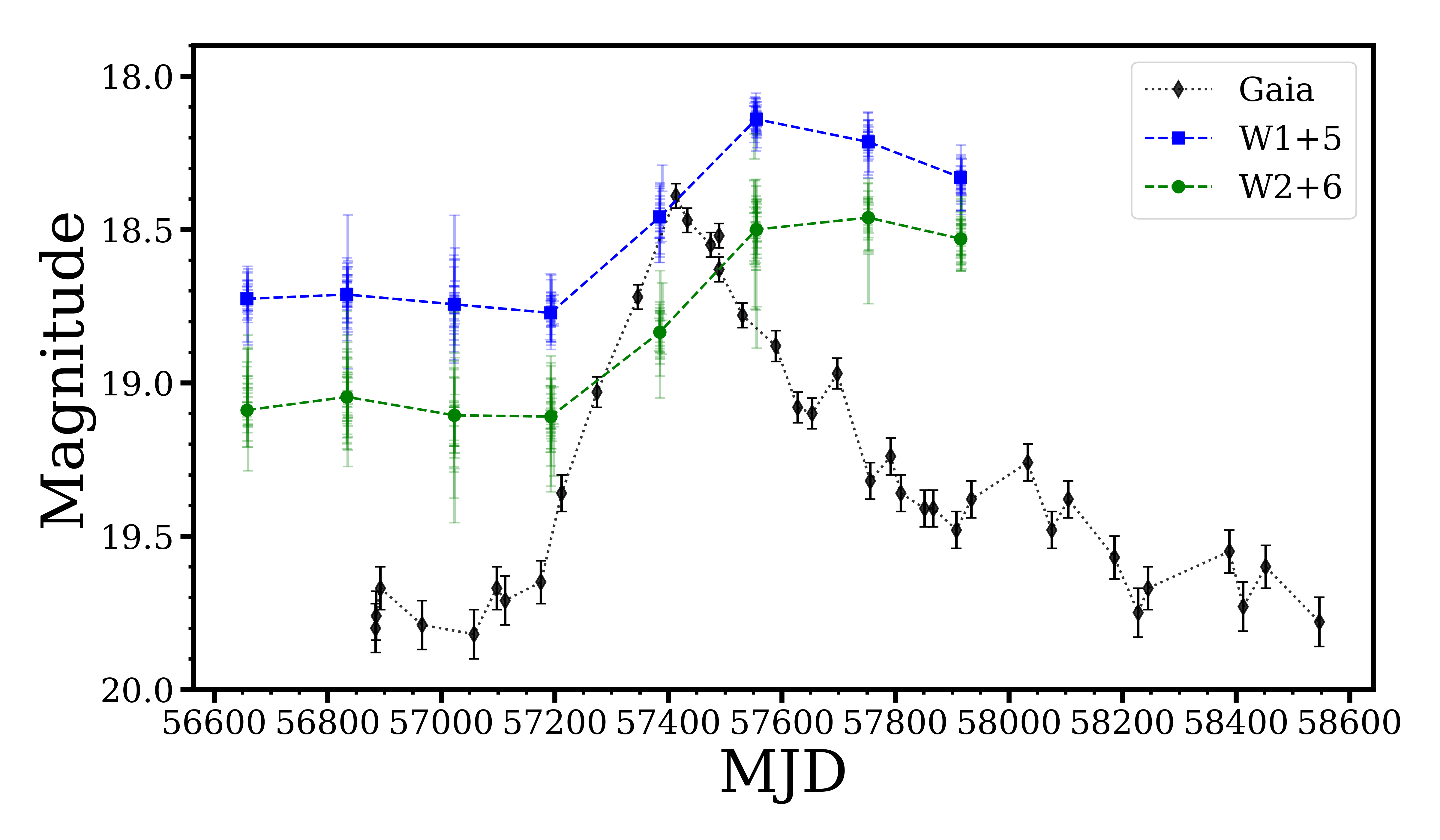}
    \caption{Lightcurve from the NEO-WISE mission, compared with the Gaia lightcurve. Plotted are the magnitudes in the two filters W1 (blue) and W2 (green), shifted along the y-axis to allow to compare with the Gaia data. Solid blue squares (W1) and green circles (W2) are the mean values of the exposures taken for each epoch. The NEO-WISE lightcurve reaches its peak with a delay of $\sim$ 140 days with respect to the Gaia lightcurve.}
    \label{fig:wise_lc}
\end{figure}

The Gaia lightcurve shown in Fig. \ref{fig:gaia_lc} shows a decay  from the peak of the emission to the pre-outburst level of emission over more than 2 years. The lightcurve shows two bumps at around 57700 and 58000 MJD (around 300 and 600 days from peak, respectively). The decay rate is well fit by an exponential decay $\sim \rm t^{-0.161\pm0.004}$ (with $\rm \chi_\nu^2=0.97$).
As shown in Figure \ref{fig:wise_lc}, where we compare the NEO-WISE lightcurve with the Gaia one, Gaia16aax reaches the peak of its MIR emission with a delay of $\sim$140 days with respect to the optical data. We assumed that the epoch at which the NEO-WISE reaches its peak to be the one corresponding to the maximum observed value (MJD $\sim$57554). As the NEO-WISE lightcurve is not as well sampled as the Gaia one, the peak of the lightcurve could have happened at a different epoch. Nonetheless, since the lightcurve evolution is smooth, we expect our estimate to be good to 100 days, which is half of the interval between consecutive WISE observations. The delay between the WISE and Gaia lightcurve is of $140\pm100$ days. Assuming that the MIR emission comes from a dusty torus surrounding the Broad Line Region (BLR), this delay would imply a distance of this torus from the central engine of $0.12\pm0.08$ pc.

\begin{figure}
    \includegraphics[width=1\columnwidth]{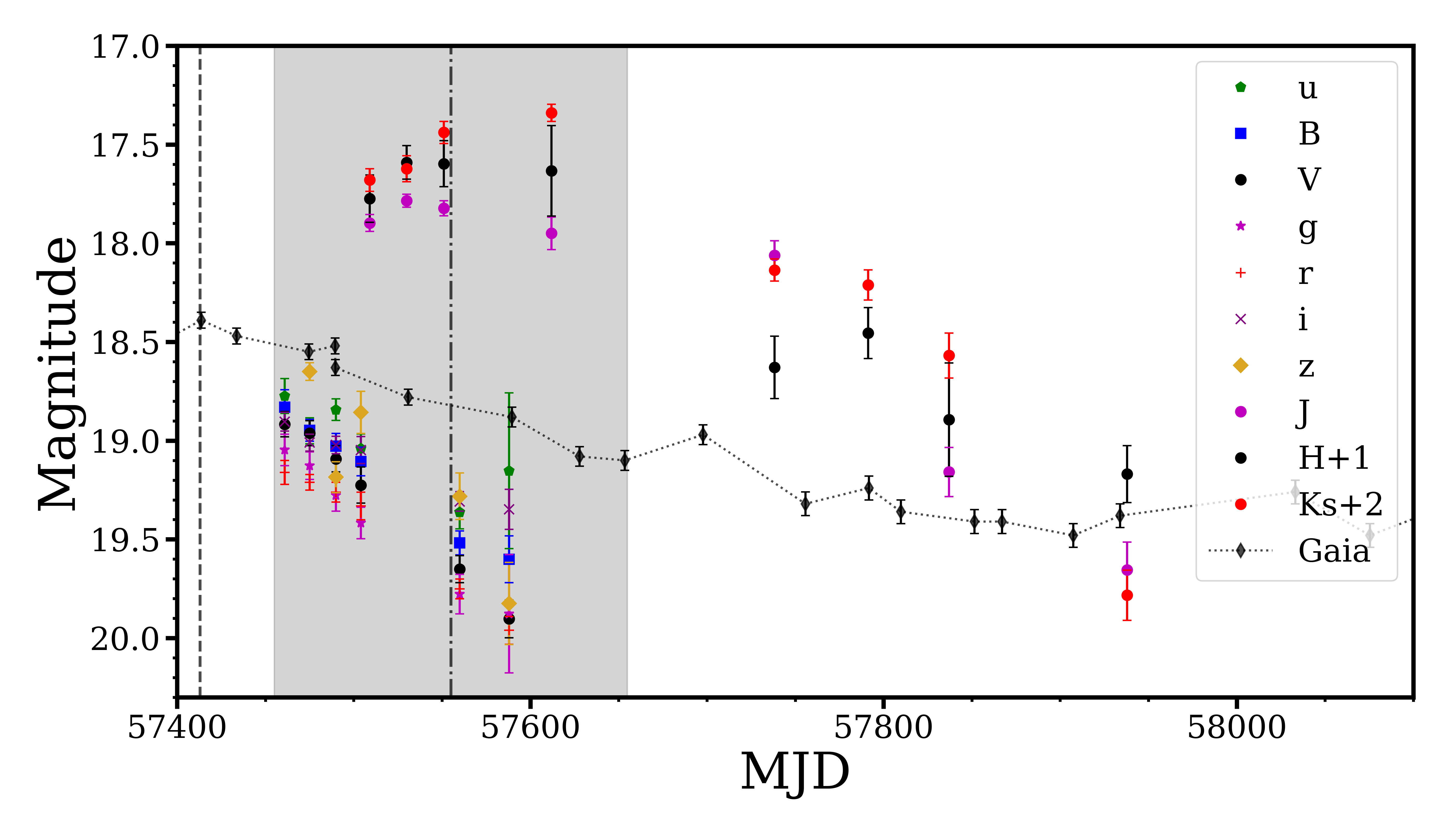}
    \caption{Light curve of the NOT and LT observations in the optical and NIR bands, compared with the Gaia lightcurve. The vertical dashed line at MJD 57\,413 corresponds to the date the alert was published while the vertical dashed line at MJD 57\,555 and the gray area indicate the time frame in which the NEO-WISE lightcurve reaches its peak. The magnitudes plotted are the ones resulting from the host-subtraction process.} 
    \label{fig:lc_opt_ir}
\end{figure}

\begin{figure}
    \includegraphics[width=1\columnwidth]{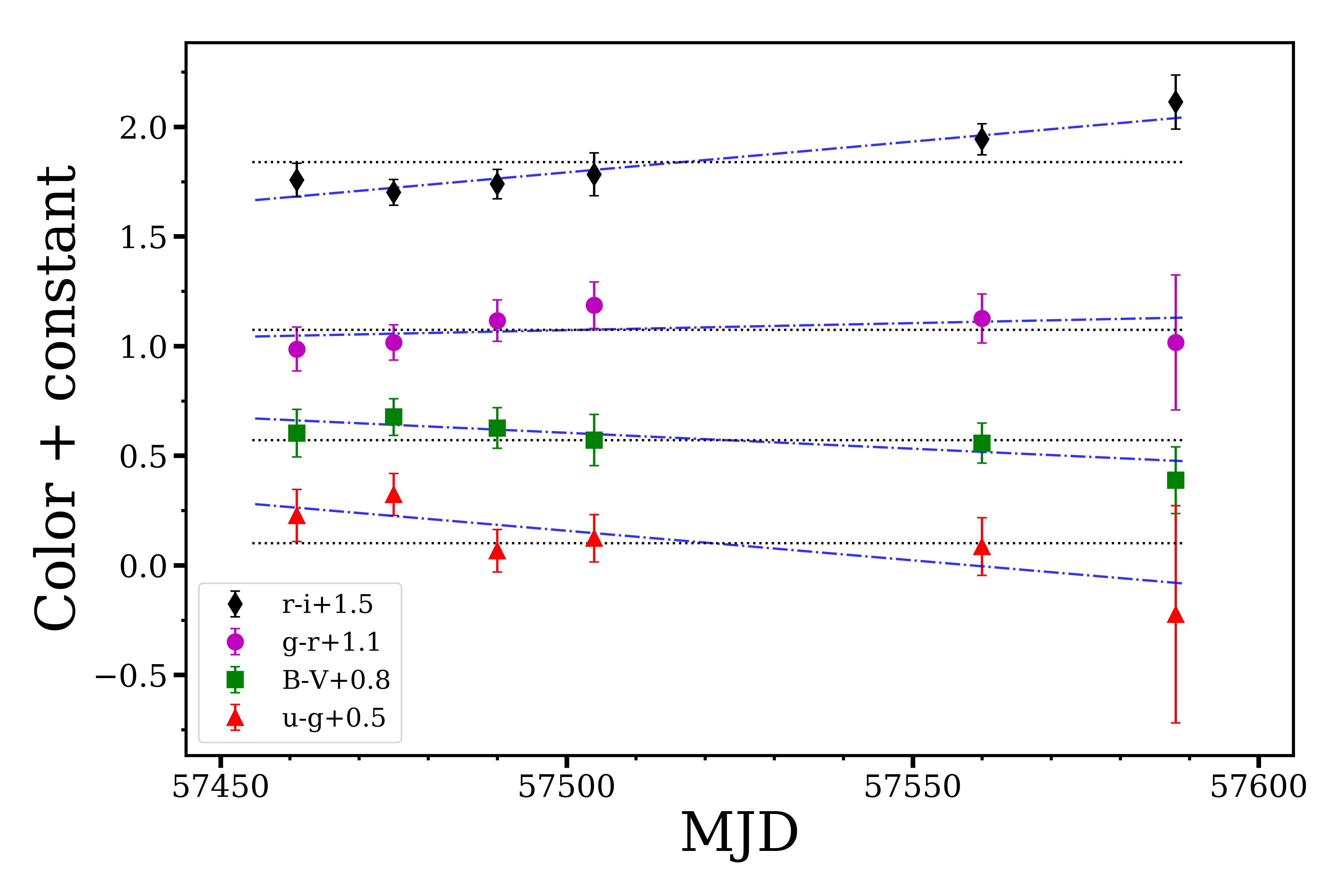}
    \caption{Color evolution of  \textit{r}-\textit{i} (black diamonds), \textit{g}-\textit{r} (magenta circles), B-V (green squares) and \textit{u}-\textit{g} (red triangles). For all the colors both a fit with a constant line (black dotted lines) and a fit with a slope (blue dashed lines). The colors are calculated from the host-subtracted magnitudes.}
    \label{fig:color}
\end{figure}

The lightcurve of NIR and optical bands is shown in Figure \ref{fig:lc_opt_ir}, and the apparent magnitudes after the host-subtraction are listed in Table \ref{tab:photometry}. The lightcurve shows a smooth decay, consistent with the behaviour observed in the Gaia lightcurve. There is a bump in the H band around 57800 MJD, this is around 100 days after the bump in the Gaia lightcurve at 57700 MJD, a value that is consistent with the delay present in the NEO-WISE lightcurve.
The color evolution is shown in Fig. \ref{fig:color}: all colors have been fitted both with a constant value and with a slope. The reduced \chisq\ in all cases is well below 1, meaning that the uncertainties on the color are too large to determine an evolutionary trend of the colors. We also performed a KS test to assess the goodness of fit and found that the colors B-V and \textit{r}-\textit{i} are better fit with a slope than with a constant. Therefore while the \textit{g}-\textit{r} and \textit{u}-\textit{g} remain constant over the period of observation, the color B-V decreases from $-0.2\pm0.1$ to $-0.4\pm0.1$  and \textit{r}-\textit{i} increases from $0.3\pm0.1$ to $0.6\pm 0.1$.

\begin{figure}
    \includegraphics[width=1\columnwidth]{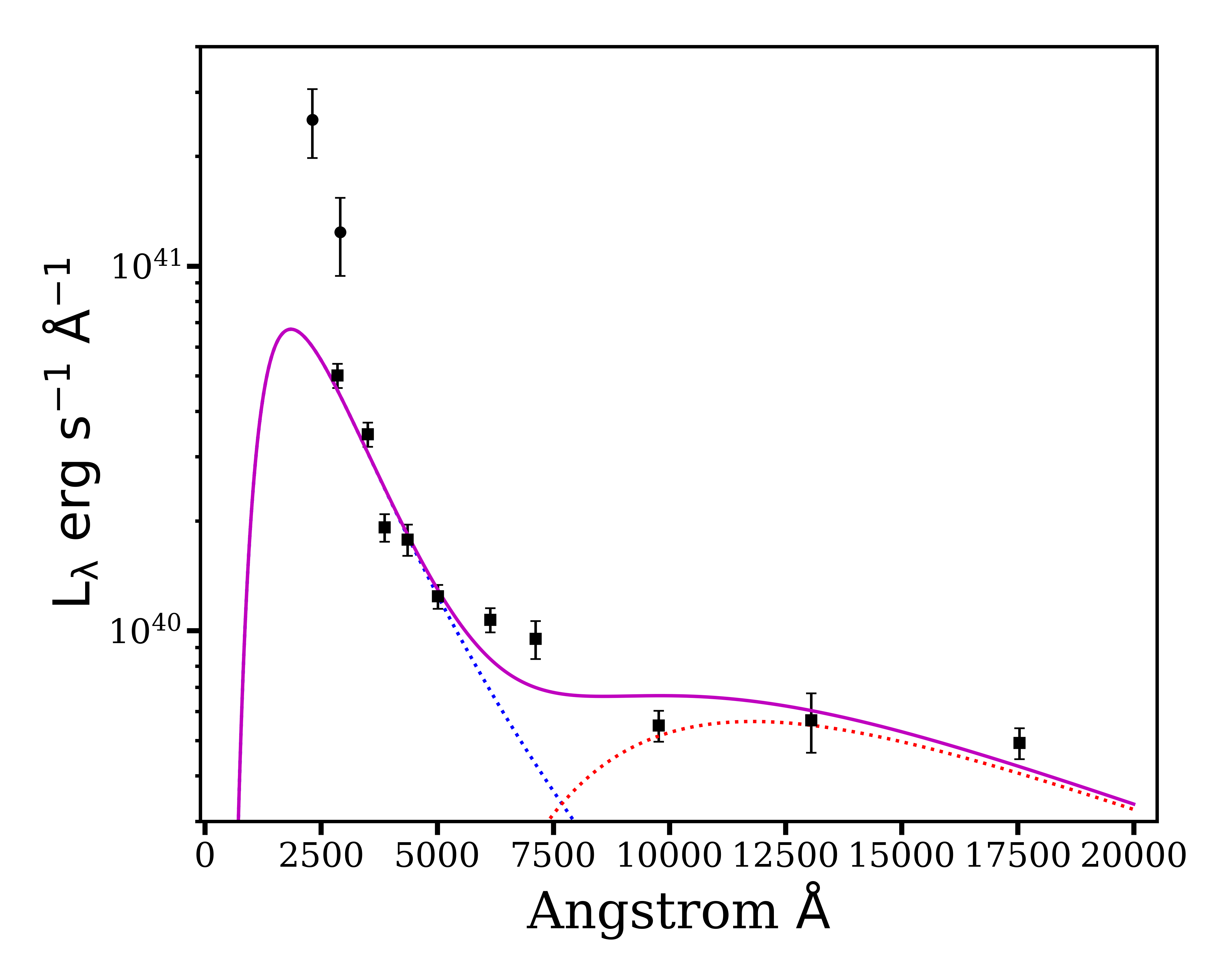}
    \caption{Fit of two black-body curves to the luminosity density. The black squares are the values of the luminosity for each filter (optical - MJD 57505 and NIR - MJD 57509), the black circles are the two values of the monochromatic luminosity for the two UV filters UVM2 and UVW1 (MJD 57569, not considered for the fit), the solid magenta line represent the fit using two black body components, these two components are represented by the blue and red dotted lines. {On the x-axis there is the rest-frame wavelength.}}
    \label{fig:bbfit}
\end{figure}

The host-subtracted and extinction corrected magnitudes were used to model the Spectral Energy Distribution (SED) and calculate the bolometric luminosity of the object using a \texttt{python} code adapted from the program \texttt{superbol} \citep{nicholl18}. For all filters the extinction correction was done using the values in \citet{schlafly11} who assume a reddening law with R$_v$=3.1. Our NIR and optical data are not coincident in time, therefore we focused on the epochs in which we had optical data, adding the NIR points to better constrain the fit where possible, extrapolating or interpolating the NIR data to match the epochs of optical imaging. However, since our first NIR observations (MJD 57\,509.04) is almost coincident with our fourth epoch of optical data (MJD 57\,505.03), we decided to add NIR data only from MJD 57\,491.01 (the third epoch of optical imaging) onward, to avoid large extrapolation of the NIR data. The total bolometric luminosity is then calculated by connecting the points in the optical wavelenghts with straight lines and integrating the area under the resulting curve, plus a black-body extrapolation in the UV and NIR regions. For every epoch we applied a K-correction to the data and then fit the fluxes derived from each band with two black-bodies. An example of the fit is shown in Figure \ref{fig:bbfit}

The absence of UV data-points hinders our ability to constrain the peak of the black-body emission. Using the UV data from our single \textit{XMM}-Newton observations we can check our choice of using two black-bodies instead of one for the optical and NIR observations closest in time to the \textit{XMM-Newton} observation: subtracting the pre-outburst UV flux measured by \textit{Swift} from the values measured from the \textit{XMM}-Newton observation for the UVM2 and UVW1 filters (central wavelength 2310~\AA~and 2910~\AA, respectively), we obtain a luminosity density of $\sim$3.2$\times$10$^{41}$erg~s$^{-1}$\AA$^{-1}$ and $\sim$1.5$\times$10$^{41}$erg~s$^{-1}$\AA$^{-1}$, respectively. These values are higher than the peak of our fit, suggesting that by using two black-bodies we are not overestimating the emission from the object.

Modeling the SED with two black-body curves does not yield a good fit (reduced \chisq $\sim4-5$ for all epochs). On the one hand, our NIR and optical magnitudes are not coincident in time, thus introducing some systematic uncertainty in our SED. On the other hand, there is probably a non-thermal component that we are not considering in our fit. This is also hinted at by the high values of the UV and X-ray luminosities. The absence of UV data, though, hinders our ability to constrain the parameters of a power-law that would probably describe the non-thermal component. On top of this, we see emission from more than one component: the optical and the (delayed) infrared emission (see Fig.\ref{fig:wise_lc}). All in all, we assume that our fit using two black-bodies is a reasonable first order description of SED and allows for an estimate of the bolometric luminosity.

\begin{figure}
    \includegraphics[width=1\columnwidth]{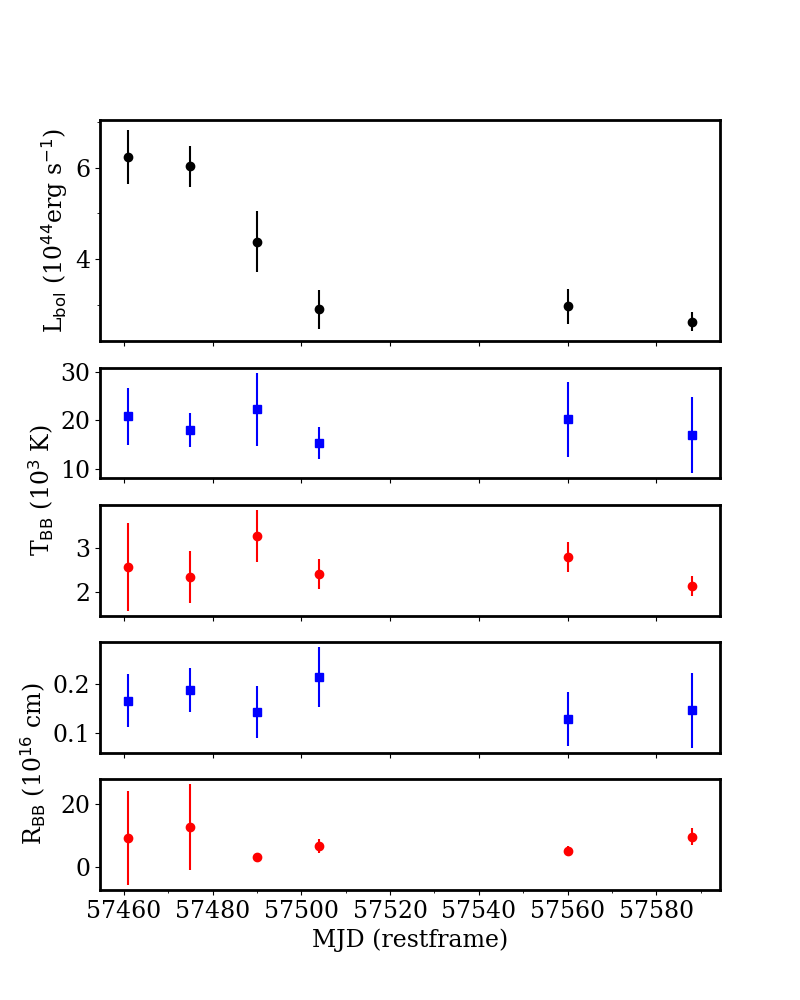}
    \caption{Results from the black-body fits to the flux density at each epoch. In the top panel we plot the bolometric luminosity, in the second and third panel we plot the temperature for the two black-bodies and in the fourth and last panel the radius for the two components (blue squares for the first black-body and red circles for the second one, in all relevant panels. The first black-body is the one that peaks at shorter wavelengths, the second one is the one that peaks at larger wavelengths, see Fig. \ref{fig:bbfit}). The large uncertainties on the black-body parameters in the first two epochs (especially for the radius of the second black-body component) are due to the absence of NIR data.}
    \label{fig:bb_results}
\end{figure}

The bolometric luminosity resulting from the black-body fits is plotted in Fig. \ref{fig:bb_results}. The bolometric luminosity of the object decreases by a factor of 2 over $\sim$300 days and then remains constant within uncertainties at around $\sim$3$\times$10$^{44}$\unitlum. The temperature and radius for both black-bodies remain constant during all the period of observation. The peak of the two black-bodies curves is at $\sim$1000~\AA~and $\sim$1 $\rm \upmu m$. The black-body temperature of the second black-body component (associated with the IR emission) is high, above 2000K, indicating the presence of dust possibly above the evaporation temperature.

We calculated the bolometric correction to the absolute magnitude from Gaia with the values of the bolometric luminosity obtained from the photometric measurements. We then calculated the average of the correction over the epochs of observations and, under the assumption that this correction remains constant, used it to estimate the bolometric luminosity from the Gaia absolute magnitudes over the duration of the whole outburst. With this method we were able to calculate the estimated total energy irradiated: $\rm E_{tot}=(3.3\pm0.9)\times 10^{52} erg$. The uncertainty on this measure has been calculated by propagating the uncertainty on the bolometric luminosity from our photometric measurements. As the Gaia magnitudes are given without uncertainty, the error on the total energy radiated during the outburst derives only from the error on the bolometric luminosity we calculated.

The bolometric luminosity evolution as a function of time calculated from the Gaia magnitudes mimics the shape of the Gaia lightcurve by design, therefore the decay trend of the bolometric luminosity is also well fit by an exponential decay $\propto \rm t^{-0.16\pm0.01}$.

We use the luminosity in the X-ray, calculated from the XMM observation, in the band 2--10 keV, $\rm L_{2-10\,keV}=(1.2\pm0.1)\times10^{44}erg/s$, to have another estimate of the bolometric luminosity. To do this, we calculate the values of the bolometric correction ($k_{bol}$) using the methods described in \citet{marconi04} and \citet{netzer09}. We obtain $k_{bol}\simeq34$, using the equation of \citet{marconi04} and $k_{bol}\simeq29.6$ from \citet{netzer09}. The resulting luminosity values are $\rm L=4.2\times10^{45}$\unitlum and $\rm L=3.6\times10^{45}$\unitlum, respectively. The uncertainties in these values are dominated by the uncertainties in the evaluation of the bolometric correction, but they are not easy to calculate. The range in the extreme values of $k_{bol}$ can be as large as an order of magnitude \citep{netzer09} and depends strongly on the constraints used to calculate $k_{bol}$.
We note that the value of the bolometric luminosity estimated from the X-ray luminosity is higher than the one calculated from the SED fit. This is in agreement with the statement that our SED fit probably underestimates the bolometric luminosity, given the absence of UV and X-ray data points.
If we, instead, compare the $L_{2-10\,keV}$ with the bolometric luminosity obtained from the SED fit, we would get a value of $k_{bol}\sim2.5$. This value is much lower than what reported in the literature \citep{brightman17,netzer09,marconi04}.
We therefore consider that the value of the total energy radiated over the outburst, obtained from the SED fit, is underestimated.

\subsection{Spectroscopic analysis}
\label{sec:spec_analisys}
We used the first spectrum (MJD 57\,428.27) to classify the transient event. This was done by cross-correlating our classification spectrum with a library of spectra using the \texttt{SNID} code \citep{blondin07}, resulting in a good match with an AGN spectrum \citep{atel16aax}. The redshift calculated from the transient spectrum is consistent with the one reported in SDSS.

In Fig.\ref{fig:gaia_spec} we plot all the spectra of our follow-up campaign. Looking at the follow-up spectra by eye, the first optical spectrum shows a blue component, while during the decay of the outburst the spectra become redder. Broad Balmer lines (\ha, \hb, \hg) are present in all spectra, as well as narrow \oiii and \nii lines. These lines are present in the SDSS quiescent spectrum, but both their intensity and structure appear to be changed in the outburst spectra. This is likely due to the difference of spectral and spatial resolutions of the spectrographs employed by SDSS and NOT. The region of the \hb~ broad emission line is complex with various broad components and narrow \oiii ($\uplambda\uplambda$ 4959 and 5007) emission lines. The \ha~line is contaminated by the telluric absorption around 8250~\AA; this hindered our ability to fit in detail the properties of the emission lines in this wavelength region. At the first epoch there is a bump in the continuum between \hg~and \hb. This is possibly due to a complex of unresolved lines such as that caused by the Bowen fluorescence mechanism \citep{bowen34,bowen35} blended with \ion{He}{ii} at 4686~\AA. Metal lines originating from the Bowen fluorescence have been recently reported to be present in TDEs \citep{leloudas19,blagorodnova19,onori19}. In the subsequent spectra, this blend is not present anymore.The absence of other strong lines typically associated with the Bowen fluorescence mechanism, such as \ion{N}{iii} $\uplambda\uplambda$ 4640, 4100 and \oiii $\uplambda$ 3760 could hint that this is instead a forest of iron lines typically seen in AGN. It is worth noting that this higher continuum is not present as distinctly as in the first spectrum at other epochs, hinting at a possible transient nature of the unresolved lines.

\begin{figure}
    \includegraphics[width=1\columnwidth]{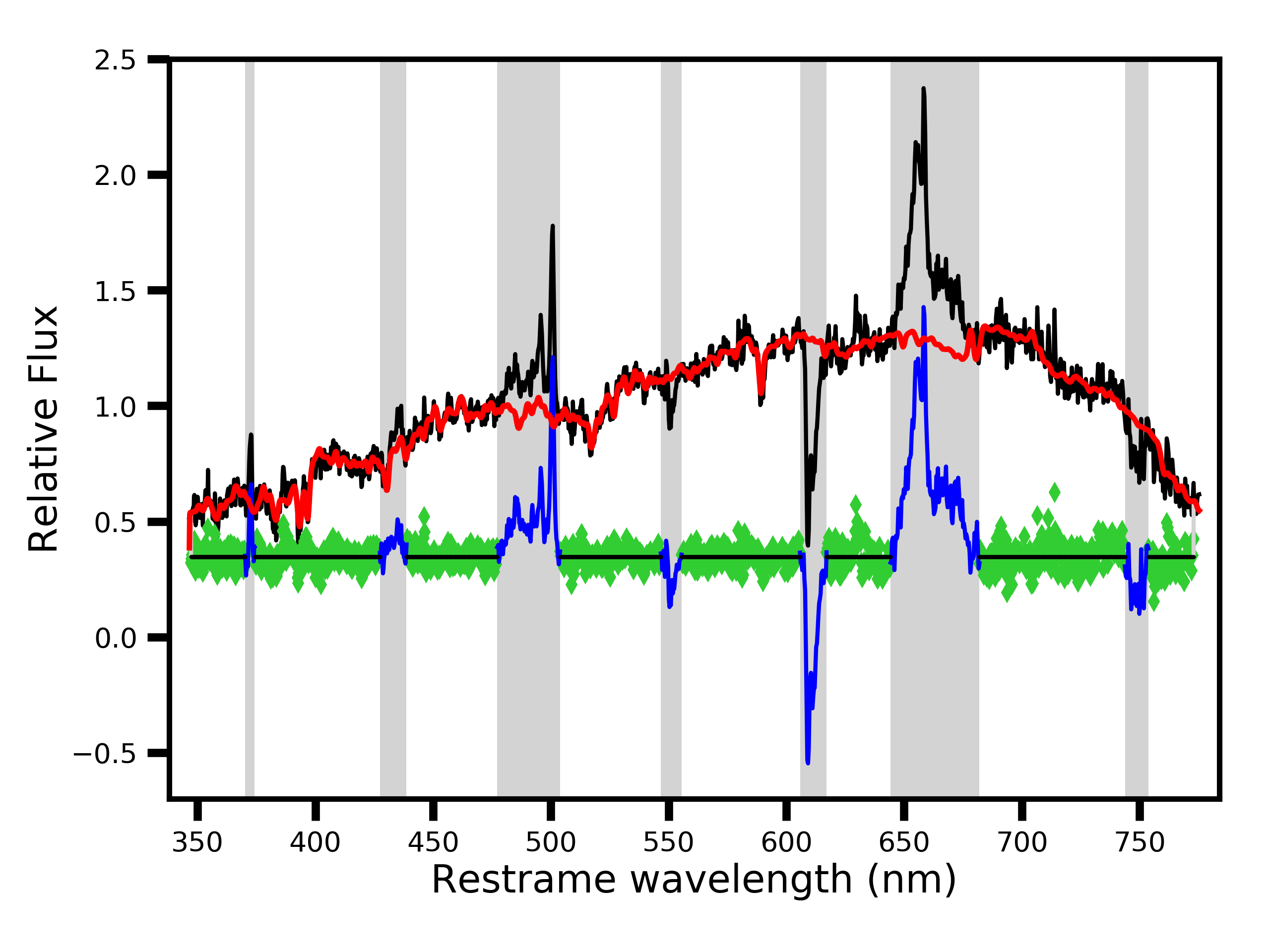}
    \caption{Fit of the stellar libraries to our last spectrum (2018 July 08, 894 days after peak) using the \texttt{pPXF} method. In black the observed spectrum, in red, the best fit of the templates from the stellar library. The grey bands represent the areas excluded from the fit. In green and blue the subtracted spectrum.}
    \label{fig:ppxf}
\end{figure}

The spectra show a significant galaxy emission component and various galaxy absorption lines, most notably the Ca H and K lines (3969 and 3934 \AA, respectively). To model the galaxy emission, we use the penalized Pixel-Fitting method \texttt{pPXF} \citep{cappellari04,cappellari17}. The method approximates the galaxy spectrum by convolving a series of N template spectra $T(x)$ with an initial guess of $f(v)$, the Line Of Sight Velocity Dispersion function (LOSVD) to the observed spectrum. The galaxy model is obtained through this parametrisation (in pixel space $x$):

\begin{equation}
\begin{split}
G_{mod} \left(x \right) = & {\textstyle \sum_{n=1}^{N}} w_n \{ \left[ T_n \left(x\right) \ast f_n \left(cx \right)\right]{\textstyle \sum_{k=1}^K} a_k \mathcal{P}_k \left(x \right)\} + \\
& {\textstyle \sum_{l=1}^L} b_1 \mathcal{P}_l \left( x \right) + {\textstyle \sum_{j=1}^J} c_j S_j \left(x \right),
\end{split}
\end{equation}

\noindent where the $w_n$ are the spectral weights, the $\mathcal{P}_k$ and $\mathcal{P}_l$ are multiplicative and additive orthogonal polynomials and $S_j$ are the sky spectra. The polynomials and sky spectra are optional components of the parametrisation. The LOSVD $f(cx) = f(v)$ is parametrised by a series of Gauss-Hermite polynomials as:

\begin{equation}
f\left(v\right) = \frac{1}{\sigma \sqrt{2\pi}} exp \left( \frac{1}{2} \frac{\left( v-V\right)^2}{\sigma^2}\right) \left[ 1 + {\textstyle \sum_{m=3}^M} h_m H_m \left( \frac{v-V}{\sigma}\right) \right],
\end{equation}

\noindent where $V$ is the mean velocity along the line of sight, $\sigma$ is the velocity dispersion, $H_m$ are the Hermite polynomials and $h_m$ their coefficients. The best fitting template is then found by $\chi^{2}$ minimisation. To compute our synthetic galactic spectrum, we chose the The Indo-US Library of Coud\'e Feed stellar spectra \citep{valdes04}, a library of 1273 stellar spectra covering a broad range of parameters (effective temperature, metallicity, surface gravity) with a wavelength range of 3460 -- 9464 \AA~and a spectral resolution of 1.35 \AA~(FWHM), $\sigma\sim$30 \kms, R$\sim$4200. We use the \texttt{pPXF} method convolving the library to our last spectrum, taken after the object went back to its pre-outburst state. We excluded from the fit the regions in which the AGN lines are present, the regions of telluric absorption and the edges of the spectrum. The templates fit is shown in Fig.\ref{fig:ppxf}. We then subtract the galaxy spectrum obtained with this method from the spectra at all other epoch. The analysis presented from here onward is performed on the galaxy-subtracted spectra.

To analyse the complex emission line structure we fit the spectra with a combination of Gaussian function using a \texttt{python} code employing the \texttt{lmfit}\footnote{https://lmfit.github.io/lmfit-py/} package \citep{lmfit}. The results of the line fits are listed in Table \ref{tab:spectroscopy}. During the outburst, except for the last two epochs, the \hb~emission line complex can be well described by two Gaussian components: component A and a red wing, component B (\hba and \hbb from here on), see Fig. \ref{fig:hb_mosaic}. \hba is blue-shifted with respect to the rest-frame wavelength of \hb~(the shift is between $\sim$5 and $\sim$20 \AA, corresponding to a velocity between $\sim$300 and $\sim$1000 \kms) while the red wing is red-shifted with respect to the restframe wavelength by a factor that varies between $\sim$50 and $\sim$80~\AA~(corresponding to a velocity between $\sim$3000 and $\sim$5000 \kms). In the \hb~region there are also two \oiii emission lines at their restframe wavelength (4959 \AA~and 5007~\AA). The separation between the \oiii lines and their FWHM have been kept fixed during the fit. 

The presence of the telluric absorption that falls on top of the \ha~region at the redshift of the object made a precise analysis of the \ha~region difficult. Overall, the \ha~emission line complex displays a morphology similar to the one of \hb~(see Fig. \ref{fig:ha_mosaic}): two components of which one (component A, \haa from here on) is blue-shifted with respect to the rest-frame wavelength (shift between $\sim$8 and $\sim$15~\AA, corresponding to a velocity between $\sim$300 and $\sim$700 \kms) and a red wing (component B, \hab from here on) which is shifted by around 100~\AA~(corresponding to a velocity of $\sim$4500 \kms). 

In addition to these two components, we fit also the two \nii lines (6550 and 6585~\AA) and [\ion{S}{ii}] (6718 and 6733~\AA, unresolved in our spectra). To reduce the numbers of free parameters in the fit of the \ha~region, we constrained the FWHM of the narrow lines to be the same as the value of \oiii~in the fit to the \hb~region at the same epoch, under the assumption that the lines coming from the Narrow Line Region do not change on the timescale of the outburst we are analysing in this manuscript and the assumption that lines of different metals have the same FWHM. The separation between the two \nii lines has also been kept fixed during the fitting procedure. On the spectrum of 2016 December 09 (MJD 57\,732) a cosmic ray hit the detector exactly on the \hba component, therefore we did not display the results from the line fitting at this epoch for both \hb~components. During the time covered by our observations, the line parameters show variations above the statistical noise, without a clear evolution with time. 

At the first two epochs (see Figures \ref{spec_ha_a} and \ref{spec_ha_b}) there is a narrow component (FWHM $\rm \sim 170$\kms) on top of the red wing of \ha, centered at $\sim$6650~\AA. This component is not present in the subsequent spectra and we associate it with either \ion{He}{i} 6678~\AA, or \ion{Fe}{i} 6648~\AA. Neither element is clearly present at other wavelengths, but \ion{He}{ii} 4686~\AA, as well as other iron lines, could be present, but there is no clear detection as the contrast with the continuum is too low.
In Fig.\ref{fig:ha_mosaic} the fits to the \ha~line in all epochs are shown. It is worth noting that the \ha~line retains the double peaked nature until our last spectrum, taken after the Gaia lightcurve reached the pre-outburst level of emission, while the \hb~is well described by only one component in the last two epochs. This could be due to the lower signal to noise ratio of these spectra and to the lower contrast between the \hb~and the continuum, with respect to the \ha~line.

\begin{figure}
    \includegraphics[width=1\columnwidth]{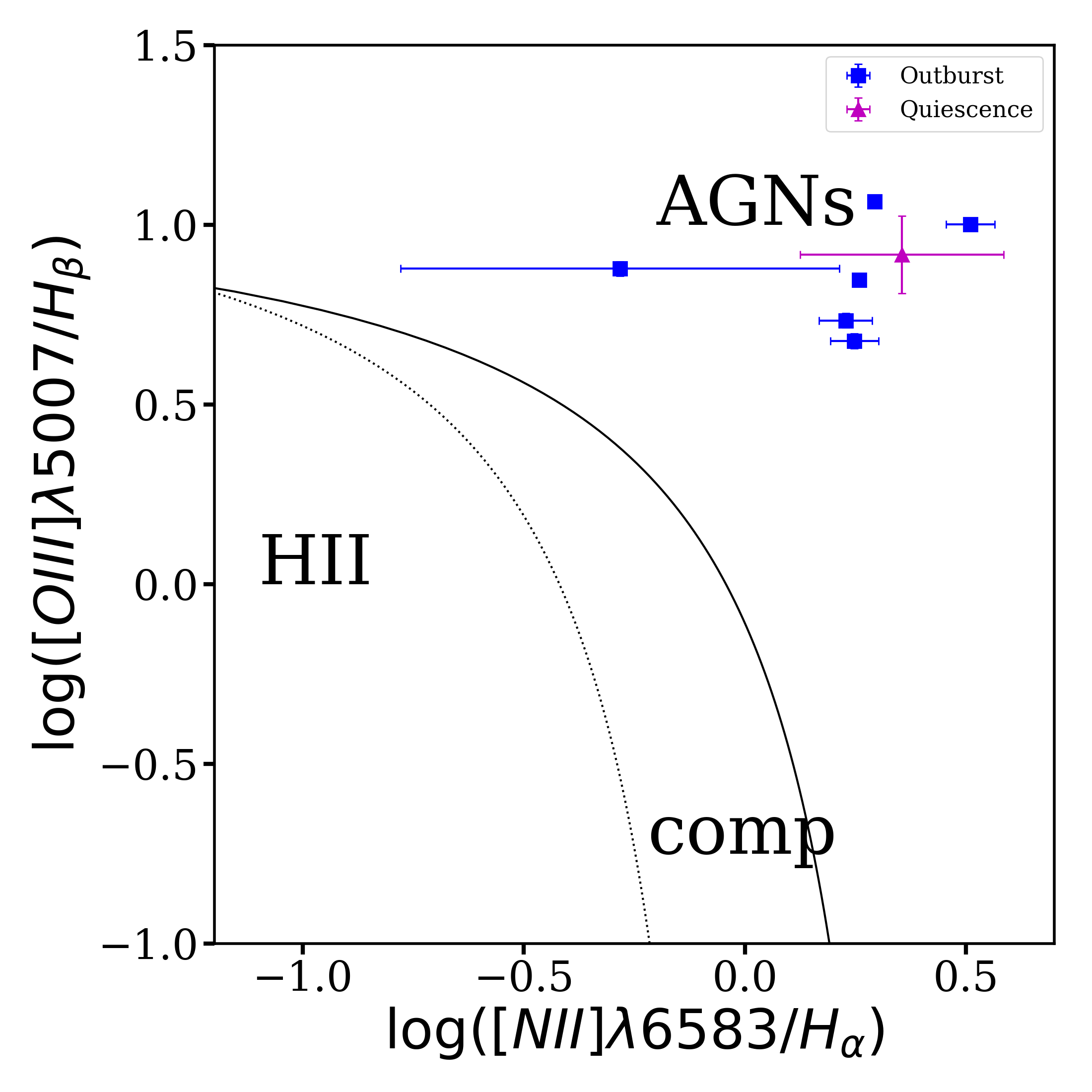}
    \caption{Plot of the position of the source on a Baldwin, Phillips \& Terlevich (BPT) diagram \citep{baldwin81} using the narrow emission lines detected in the host-spectrum of Gaia16aax, during the outburst (blue squares) and after it went back to the pre-outburst state (magenta triangle). In all cases the position is consistent with the AGN region of the diagram}. The lines that separate the different activity regions come from \citet{kewley01} (dotted line) and \citet{kauffmann03} (solid line).
    \label{fig:bpt}
\end{figure}

In Figure \ref{fig:bpt} we show the position of the source in a Baldwin, Phillips \& Terlevich (BPT) diagram \citep{baldwin81}. For the calculation of the ratios of the fluxes we introduced a narrow \ha~and \hb~component. To do this, we first fitted this additional component in the first spectrum (the one with the highest signal to noise ratio), constraining its FWHM to be the same of the other narrow lines at the same epoch. In the subsequent spectra, we kept the parameters of the narrow \ha~and \hb~component fixed, within uncertainties, under the assumption that the narrow lines do not change over the timescale of our follow-up. We measured the ratios for the NOT spectra taken during the outburst decay and our last spectrum taken when the source was back in its pre-outburst state. At all epochs in which the narrow \ha~ and \hb~ components could be constrained, the position in the BPT diagram is consistent with the AGN area of the diagram, meaning that the NRL is dominated by AGN ionisation.

\subsection{Black hole mass estimate}
We use the  $M-\sigma_*$ relation \citep{ferrarese05} to obtain an estimate of the BH mass of Gaia16aax, using the width of the Ca H+K lines to estimate the velocity dispersion of the galaxy. For this calculation, we use the second spectrum of our follow-up, as it is the one with the highest signal to noise ratio (SNR), after the first classification spectrum, but less contaminated by the outburst light. After correcting for the instrumental broadening (16.2 \AA~for a 1.0$\arcsec$ slit), we calculate $\upsigma = 264\pm58$\kms and a BH mass of $M_{bh}=(6.4\pm3.7)\times10^8M_\odot$. The uncertainty on this measure is large, considering the modest SNR of the spectrum and the resolution of the instrument. The scatter in the  $M-\sigma_*$ relation, which is 0.34 dex \citep{ferrarese05}, contributes to the uncertainty of this mass estimate. The resulting Eddington luminosity is $L_{Edd}=(7.9\pm4.5)\times10^{46}erg/s$.\\
We are able to use the same method also for Gaia16ajq, where the Ca H+K lines are visible in the outburst spectrum. We obtain a BH mass of $M_{bh}\sim4.5\times10^8M_\odot$.
In the case of Gaia16aka, the Ca H+K lines are not visible, therefore we use two Single Epoch (SE) scaling relations between the continuum luminosity, the width of the \ha~\citep{greene10} or \hb~\citep{vestergaard06} line and the BH mass. To obtain the continuum luminosity of the AGN and the width of the line, we first subtract the galactic component from the spectrum of Gaia16aka, using the \texttt{pPPXF} procedure described before. We then fit the subtracted spectrum with multiple Gaussian components. From the SE relation described in \citet{greene10}, using the width of the \ha~line, we obtain a BH mass of $M_{bh}\sim2.4\times10^7M_\odot$, while from the relation between the BH mass and the width of the \hb~line we obtain $M_{bh}\sim3.7\times10^7M_\odot$. It is important to notice that these single epoch relations are intrinsically uncertain and may only yield an order of magnitude estimate \citep{vestergaard06}.

\section{Discussion}
\label{sec:discussion}
Before discussing in detail the possible interpretations of Gaia16aax, it is worth summarising the main properties of the event:
the transient is coincident with the nucleus of a galaxy that hosts a radio-quiet QSO ($\rm \sim0.02\arcsec$ separation), with an inferred \mbh~of $\rm \sim6\times10^8$\Msun. The source brightened in the optical by 1 magnitude over a timescale of $\sim$350 days, reaching a peak absolute magnitude of $\rm M_G$=-22.17 and started a smooth decay with $\rm t^{-0.161\pm0.004}$, going back to its pre-outburst level of emission over 2 years. Flaring activity of this magnitude was detected both in the MIR (with a time lag of $\sim$140 days) and in the X-rays. The first spectra showed a strong blue continuum, while in the subsequent epochs the object became redder. In the first spectrum a blend of lines between \hg~and \hb~is present, this could be due to a blend of \ion{He}{ii} and Bowen fluorescence emission lines. During the outburst, the Balmer lines underwent a dramatic change in their morphology, showing a clearly separated double-peaked profile. The narrow lines did not change over the period of our observations. In the first epochs of observations the spectra showed a strong blue continuum that disappeared with time. The red wings of both the \ha~and \hb~lines are offset from their rest-frame wavelength by several thousand \kms and this shift does not vary significantly in time. The width of the lines and their equivalent widths do not show significant evolution with time. The SED cannot be satisfactorily fit using a single black body, and a reasonable fit, although not perfect, is obtained using two black body components.


\subsection{AGN variability}
\label{sec:agn_var}
The enhanced emission and change of appearance of the emission lines could have been caused by variability in the accretion disk. While typical AGN variability is of tenths of magnitudes over timescales of some years \citep{vanvelzen10}, more extreme cases of variability have been found. In \citet{rumbaugh18} a big sample ($\sim$1000 objects) of Extremely Variable Quasars (EVQs) has been studied, finding that Quasars that show extreme variability (i.e. a change of the optical magnitude >1 mag) all have a low Eddington-ratio (L/L$\rm_{Edd}$<0.3). They argued that EVQs are not a separate class of objects, but rather a subset of quasars that are accreting at low rates. While individual events may originate from peculiar phenomena, the majority of EVQs are thought to be caused by accretion related events.
Gaia16aax throughout all its outburst maintains an Eddington-ratio below 0.3, in line with the sample of EVQs presented in \cite{rumbaugh18}. However, the rapid time scale of Gaia16aax sets it apart from the majority of the objects in \citealt{rumbaugh18} sample. In their sample, there are around 50 objects ouf of 977 that reach the maximum amplitude variability in a time $\leq400$ days.

\subsubsection{Timescale and disks}

It is of interest to compare the timescales found in Gaia16aax, (i.e. a rise timescale of hundreds of days and a decay timescale of years) with the characteristic timescales associated with different aspects of accretion disks in AGN. 

\citet{lawrence18} has recently refocused attention on a long-standing problem (see also, \citealt{antonucci18}) that quasar variability is not compatible with standard accretion disk theory. The problems are succinctly summarised by \citet{dexter19}. In agreement with Lawrence's conclusion that "the optical output we see comes entirely from reprocessing of a central source", \citet{kynoch19} argue that the most favoured explanation of the observations is that the variability at all wavelengths can be accounted for by an intrinsic change in the luminosity of the central object (and the brightening observed in Gaia16aax in UV and X-rays seems to confirm this) -- presumably the central regions of the black hole accretion disk. Assuming that this is so, then it appears that the most severe problem is the timescale of the variability (see also \citealt{stern18}). 

Such large amplitude variations in luminosity must be caused by variations in the accretion rate in the inner accretion disk (e.g. within $R < 10 - 20 R_g$, where $R_g = GM/c^2$ is the gravitational radius). At such radii for the more luminous quasars, standard accretion disk theory \citep{shakura73} finds that the disks are radiation pressure dominated. In that case the standard viscous timescale for radial inflow $t_\nu \approx R^2/\nu$, where $\nu$ is the effective kinematic viscosity, can be written (e.g. \citealt{pringle81})
\begin{equation}
t_\nu \approx (\Omega \alpha)^{-1} (H/R)^2,
\end{equation}
where $\Omega$ is the local disk angular velocity, and $H/R$ the disk aspect ratio,
or, equivalently, in this case (cf. \citealt{dexter19})
\begin{equation}
\label{tnu}
t_\nu \approx 43 \left( \frac{\alpha}{0.3} \right)^{-1} \left( \frac{\kappa}{\kappa_T} \right)^{-2} \left( \frac{M}{10^8 M_\odot} \right) \left( \frac{\dot{m}}{0.1} \right)^{-2} \left( \frac{R}{10 R_g} \right)^{7/2} \; {\rm d}.
\end{equation}

Here $\alpha$ is the usual viscosity parameter, and we have adopted the observed value of $\alpha \approx 0.3$ for fully ionised disks \citep{martin19,king07}, $\kappa$ is the opacity, $\kappa_T$ the electron scattering opacity, $M$ the black hole mass, and $\dot m = 0.1 {\dot M} c^2/L_{\rm Edd}$ is the dimensionless accretion rate, where $\dot{m} = 1$ gives a luminosity at the Eddington limit, $L_{\rm Edd}$ for an assumed radiative efficiency of $\epsilon = 0.1$.

Thus we see that while the inflow timescale can be as short as months at very small radii, the timescale depends strongly on radius. Thus, for example, in order to account for changing look behaviour in which to whole of the inner disk regions (say, out to $R \approx 50 - 100 R_g$) need to be removed and/or added would require, according to Equation (\ref{tnu}) a timescale of some 30 -- 370 years.

\subsubsection{Cause of the variability}

In addition to the timescale problem with standard disks, there is a more fundamental problem, which is that the reason for such disks to be variable at all has not been identified. Most of the ideas on what might cause fluctuations is the disk mass flow centre on local variability within the disk. \citet{kelly09} and \citet{dexter11} propose {\it ad hoc} thermal fluctuations within the disk (see also, \citealt{nowak95,ruan14}). \citet{ingram11} propose {\it ad hoc} local fluctuations in the accretion rate, $\dot M$. More physically, fluctuations in local magnetic processes have been proposed by \citet{poutanen99} (flares in the corona), \citet{hawley01}, (instabilities in the plunging region of the inner disk) \citet{hogg16}, (fluctuations caused by local dynamo processes) and \citet{riols16} (cyclic dynamo and wind activity). As pointed out by \citet{king04}, all these ideas have the same fundamental drawbacks: the timescales for the fluctuations are too short (typically a few local dynamical timescales), in that they are much less than the local inflow timescale and so do not propagate far radially. Thus, being essentially localised, they all produce low amplitude fluctuations.

\subsubsection{Large amplitude variability}

The problem of producing large amplitude fluctuations in the luminosity of the disk requires a large amplitude fluctuation at small radii (where most of the accretion energy is released) but on a timescale much longer than the dynamical timescales at those radii (typically hours to days). This problem has been addressed by \citet{king04}, though in a different context. King et al. drew on an earlier idea by \citet{lyubarski97} who pointed out that if some mechanism could be found for varying the accretion rate  at large radius, {\it on the timescale for inflow at that large radius}, then the resulting accretion rate fluctuation would be able to propagate to small radii, and so produce a large amplitude luminosity fluctuation on that timescale. This is the underlying basis of the {\it ad hoc} fluctuation analysis of \citet{ingram11,ingram12}.

\citet{king04} proposed a physical model which was able to generate accretion rate fluctuations in the disk on timescales much longer than the local dynamical timescale. The basis of the proposal is that from time to time at least some of the angular momentum of disk material is removed by a magnetic wind (cf. \citealt{blandford82}) and not just by disk viscosity. In order for such a wind to be effective at a given radius, it is necessary that there is a strong enough poloidal field component threading the disk at that radius. \citet{lubow94a} demonstrated that such a field cannot be transported inwards by the accretion flow itself, because, for an effective magnetic Prandtl number of unity (likely for Magneto Hydro-Dynamic turbulence), the field can move radially through the disk on a timescale $t_B \approx (H/R) t_\nu$, that is with a speed $v_B \sim (R/H) v_R$, where $v_R \approx \nu/R$ is the usual viscous flow speed. Thus it is more likely that such a field, if it exists, must be generated locally by local MHD processes involving an inverse cascade in order to produce a field with spatial scale $\sim R$ necessary for a wind, from the dynamo disk scale $\sim H$. That such a process can occur in disks has been suggested by \citet{tout96} and \citet{uzdensky08}.

\citet{lubow94b} pointed out that such a locally-driven wind can in principle produce an inflow velocity that is faster than the usual viscous flow speed $v_R$.  They proposed that if the effect of the wind became locally strong enough that the wind-driven inflow speed exceeded $v_B$, then the poloidal field responsible for the wind could be dragged inwards. In that case, they suggested, there is the possibility for a wind-driven avalanche which could sweep the inner disk regions inwards. \citet{Cao02} and \citet{campbell09} concur with this conclusion. An example of how such an avalanche might operate is to be found in the numerical simulations by \citet{lovelace94}.

From the point of view of timescales, this mechanism (if and when it occurs) has the advantage that it occurs on a timescale shorter than the usual viscous timescale (Equation \ref{tnu}) by a factor of $H/R$. that is, on a timescale
\begin{equation}
\label{tB}
t_B \approx 6 \left( \frac{\alpha}{0.3} \right)^{-1} \left( \frac{\kappa}{\kappa_T} \right)^{-1} \left( \frac{M}{10^8 M_\odot} \right) \left( \frac{\dot{m}}{0.1} \right)^{-1} \left( \frac{R}{10 R_g} \right)^{5/2} \; {\rm d}.
\end{equation}
At a radius of $R \approx 50 R_g$ this corresponds to a time-scale of about a year [335 days].

What remain, of course, very uncertain are the time-scales on which such avalanche episodes might recur, and how these might depend on specific disk properties. \citet{king04} proposed a specific (speculative and, of course, {\it ad hoc}) model, which has some physical basis. They propose the basic idea that each disk annulus of width $\sim H$ acts as an independent producer of vertical field $B_z$ which varies stochastically on the local dynamo timescale $\sim 10 - 20 \; \Omega^{-1}$. They suggest, further, that occasionally enough ($\sim R/H$) neighbouring annuli have $B_z$ aligned to produce a poloidal field with length-scale $\sim R$, and that when that happens angular momentum can be lost to a locally produced magnetically driven wind. They find that most of the time such processes give rise to frequent small amplitude luminosity fluctuations, but they do give one example of one such large-scale fluctuation which began at a few hundred $R_g$, which swept rapidly inwards, reducing the overall disk surface density by almost an order of magnitude.


\subsection{Supernova}
To comprehensively survey all other plausible scenarios, we also consider if the observed outburst could be explained by a supernova explosion in the nuclear region of the galaxy. A supernova explosion in the proximity of the nucleus (or superimposed spatially on our line of sight) would not explain the shape change of the Balmer lines. On top of this, the absolute magnitude of the peak of the outburst ($\rm M_G\simeq-22$) is uncomfortably high to be caused by a SN and SNe are rarely seen emitting at X-ray wavelengths \citep{dwarkadas12} and would not explain the observed enhanced X-ray luminosity. Finally, even with efficient conversion of kinetic energy to radiation, the total radiated energy of Gaia16aax strains most reasonable supernova scenarios. Nonetheless, It is worth noting that \citet{kankare17} discuss a SN origin for the nuclear transient PS1-10adi, that happened in the nucleus of a galaxy hosting an AGN and radiated a total energy of $\sim2.3\times10^{52}$ erg.

\subsection{Microlensing Event}
Microlensing events have been invoked to explain highly variable AGNs: in \citet{graham16}, 9 of the 51 objects analysed are well described by a single-lens model and \citet{lawrence16} have proposed that microlensing provides a good description for many of the objects in their sample. 

A microlensing event could in principle explain the intensity of the outburst, but it would not be possible to explain the change in shape of the Hydrogen lines and their time evolution via this phenomenon alone. On top of this, if a microlensing event would indeed be the cause of the enhanced emission, we would expect the lightcurve to be symmetric, while in our case the decay is much shallower than the rise to peak. The presence of a secondary bump at late times in the Gaia lightcurve also disfavours a microlensing event.


\subsection{Variable dust absorption}
For completeness, we also investigate whether the change of appearance in Gaia16aax could be due to variable absorption in our line of sight.
In the case of Gaia16aax it is easy to see why this is not a viable explanation with: using the method described in \citet{macleod16} from the monochromatic luminosity of the QSO at 5100 \AA, we can estimate the size of the BLR to be R$\rm _{BLR}\sim$10 light days (with $\rm \lambda L_\lambda(5100\AA)\simeq1\times10^{43}$\unitlum, from the SDSS spectrum) or $\rm \sim2.6\times10^{16}~cm\sim$0.01 pc. We need to draw this curtain in the line of sight on a timescale of $\sim$100 days (the rise time of the outburst), that means that the cloud must travel at (10/100)c or $v\simeq3\times10^4$ km/s. To obtain the required reddening of A$_{\rm v}\simeq$1 mag we would need a column density (assuming standard dust to gas ratio) $\rm N_H\simeq 2 \times~10^{21}~cm^{-2}$, assuming a spherical cloud this would lead to a volume density $n\simeq \rm N_H/R_{BLR}\simeq 2.3\times10^4~cm^{-3}$. Such an object would correspond to a dense core within a molecular cloud \citep{blitz99}. Assuming that the object is in virial equilibrium, it must have an internal dispersion velocity of $\rm \sigma\simeq(GM_{cloud}R_{BLR}^{-1})^{1/2}\simeq(Gm_pN_HR_{BLR})^{1/2}\simeq30~km/s$. The difficulty of this scenario would be to have this cloud move at the required speed of $\sim$ 0.1 c.

If the cloud was closer to the observer, the requirement on the speed would be less stringent. Assuming that the cloud is at a distance d=$f\rm D_L$, with 0 < $f$ < 1 and $\rm D_L$ the luminosity distance (1.26 Gpc), then the required size of the cloud is R=$f\rm R_{BLR}$ and the required velocity is $v=3f\times10^4$ \kms.
The column density $\rm N_H$ must be kept fixed because of the required reddening value, therefore the volume density (assuming a spherical cloud) must increase accordingly: $n\simeq 2.3f^{-1}\times10^4~\rm cm^{-3}$. To maintain the cloud internal virial support we will have $\sigma\propto f^{1/2}$.

If we consider a cloud in the outskirts of an intervening galaxy with a velocity of $v\sim100$ \kms, we would need $f=3\times10^{-3}$ and a cloud of size $\rm R=7.8\times10^{13}cm\simeq6\,AU$ and a density of $n\simeq8\times10^6\rm cm^{-3}$.
Considering instead a cloud in the local group we would need $f=10^{-4}$ to have $\rm d=126$ kpc. Then we would have $v\simeq3$\kms and $n\simeq2.3\times10^8\rm cm^{-3}$. If the cloud is instead inside the Milky Way, we would consider $f=10^{-5}$, $\rm d=12.6$ kpc and $n\simeq2.3\times10^9\rm cm^{-3}$. In all these cases the requirements on the velocity and/or the volume density of the intervening clouds make this scenario improbable.
The fact that the outburst was observed with similar amplitude in the NIR and X-rays also plays against an absorption scenario: we would expect some form of reprocessing of the radiation at different wavelengths.
In addition to this, the source is undergoing an outburst rather than a dimming episode and it is the first time it has been observed in such a state. This would mean that we would need a cloud with the described properties constantly obscuring the central engine, until a 'hole' with the right density gradient would expose the source to us.
On top of this, it is unclear how a variable absorption scenario would explain the appearance of the double peaks in the \ha~and \hb~emission lines.
The delay observed between the peak of the Gaia and WISE lightcurves also disfavours this variable absorption scenario. In fact, if we suppose that the enhanced emission comes from de-obscuration rather than an accretion-related event, we would not expect the IR emission to change similarly to the optical emission. This is because in this scenario, the IR emission would emerge, in an isotropic fashion, from the dust heated by the central engine. If the central emission is not varying, even if the obscuration in our line of sight is changing, we would expect the IR emission from the dust to remain unchanged.


\subsection{Tidal Disruption Event}
Tidal disruption events typically show a fast rise to a peak luminosity around $10^{44}$ \unitlum, followed by a decay that follows $t^{-5/3}$. The decay time of our outburst is too long to fall into the canonical picture of TDEs, but there have been examples of long TDEs, the most extreme cases being a decade-long TDE candidate, see \citet{lin17} and a long lived TDE in a merging galaxy pair Arp 299 \citep{mattila18}.
TDEs are known to show broad (up to 10\,000 \kms) Hydrogen and/or Helium lines. The absence of color and black body temperature evolution of Gaia16aax, albeit on the limited time span of our photometric and SED coverage, is  compatible with a TDE scenario \citep{arcavi14}.

The high mass inferred for the SMBH, though, complicates the TDE scenario. In fact, a sun-like star will be swallowed whole by the SMBH if its mass is above $10^8\rm M_\odot$. To explain  Gaia16aax as a TDE, the disrupted star must be more massive and/or the SMBH must be rapidly spinning (\citealt{hills78}, see also the TDE candidate ASASSN-15lh, \citealt{leloudas16}). For a maximally spinning BH, the limit on the BH mass for a TDE to take place can increase by an order of magnitude \citep{kesden12}. 
The high spin, on top of the high BH mass, introduces new arguments in favour of a TDE interpretation. Depending on the BH spin and if the to-be-destroyed star's orbit is prograde or retrograde, the radiation efficiency can vary by a factor of 10 \citep{bardeen72}, therefore the presence of a spinning BH could justify the high energy output of the event. On top of this, simulations performed by \citet{guillochon15} found that for massive BHs ($\gtrsim10^7M_\odot$) the strong general relativistic effects produce a rapid circularisation of the debris giving rise to a prompt flare, explaining the fast rise to peak of Gaia16aax.

If the BH is maximally spinning, the inner disk temperature will be higher than in the case of a non rotating BH. We tried to constrain the upper temperature of the hotter of the two black bodies that describe the SED by fitting the combined UV and optical magnitudes (up to $\sim$5000\AA~in Fig. \ref{fig:bbfit}). Since the temperature depends on the size of the Innermost Stable Circular Orbit (ISCO, which is smaller for spinning BHs than for non spinning ones), by constraining the maximum temperature of the hotter black body, we could constrain the size of the ISCO. The observed data are not compatible with the tail of a hot ($\sim10^7$ K) black body, as the emitting region associated with such a black body has a size of $\sim10^5r_g$ where $r_g$ is the gravitational radius. Therefore we cannot constrain the spin of the SMBH.

TDEs are usually found and studied in inactive galaxies, although one of the canonical TDEs ASASSN~14li occurred in a low-luminosity AGN \citep{prieto16}. The theoretical predictions and observational properties of TDEs in galaxies that harbor an AGN are less well-constrained. In \citet{chan19} it is shown that a stream of debris from a disrupted star will impact the accretion disk and drain the accretion disk from the point of impact inwards on timescales shorter than the inflow time from a \citet{shakura73} disk, with timescales that depend on the orbital parameters of the disrupted star. The disk will then replenish itself on the (longer) inflow timescale. These two different timescales could explain the difference between the observed rise and decay times. This scenario could also help to explain the shape of the \ha~and \hb~emission lines and the secondary peaks we observe in the light curve decay. Initially, the debris stream coming from the disrupted star slams into the accretion disk, starting the drainage of the disk material, resulting in the enhanced luminosity. Due to the impact, some of the material will splash out at an angle. If we consider this outflowing material as part of the source of the \ha~and \hb~emission lines (the other one being the BLR), the inclination of the outflow angle with respect to our line of sight would explain the double peaked shape of the lines. If the initial debris stream is dense enough, the stream will pierce through the accretion disk, and slam into the disk again at a subsequent passage. This multiple encounters between the stream and the disk would help explain the presence of the two bumps in the light curve decay at $\sim$300 and $\sim$600 days from peak.\\
\indent It is worth noting that in the case of a TDE interacting with an AGN disk, the interaction between the disrupted material and the accretion disk - and thus the observed properties - will highly depend on the angle of incidence. One could argue that the discovery of Gaia16ajq and Gaia16aka could play against a TDE interpretation, given the almost identical light curves and spectra of the three objects. Nonetheless, there are differences between the three objects that could reflect a diversity of encounters between the tidal debris and the accretion disk. Only Gaia16aax shows two components in the \ha~and \hb~emission lines, which in our picture are associated with splashing material after the encounter between the debris stream and the accretion disk. Moreover, the light curve of Gaia16aka decays smoothly, without secondary peaks, while Gaia16ajq shows a very strong bump $\sim400$ days after peak. As stated before, the bumps in the light curve are associated with eventual multiple interactions between the debris stream and the accretion disk.\\
\indent If the three events are due to TDEs around spinning SMBH, Fig. \ref{fig:gaia16_lc} requires the characteristic timescales involved (e.g. mass return time and cooling time) as well as the total energy output to be very similar in each case. The SMBH masses in these sources vary by a factor of $\sim 10$, implying cooling and mass return times that vary by a factor $\sim 2-3$ for similar stellar mass and pericenter (see \citet{chan19} for the relevant equations). For the lightcurves to lie on top of each other as in Fig. \ref{fig:gaia16_lc}, we would need fine tuning between the stellar mass and radius, the SMBH mass and the pericenter distance of the encounter. It is also possible that the Gaia selection criteria mean that we are biased towards flares of this form. Implying that there could be a large number of flares due to TDEs in AGN that are not being detected efficiently by the Gaia Alerts system (e.g.~\citealt{kostrzewa18}).\\
\indent We also note that \citet{chan19} suggest that parabolic TDEs should cause a dip in the AGN X-ray flux as the corona is obscured by the tidal tail. In Gaia16aax, the X-ray emission increases by almost one order of magnitude with respect to the pre-outburst value. If the X-ray flux observed is associated with the TDE, it implies the TDE may have occurred close to the plane of the AGN disk. 
The fact that the object is still bright in the UV and X-rays years after the peak is also in line with what seen in other TDEs (e.g.~\citet{brown17,jonker19}).\\
\indent It is interesting to consider also the case of the TDE Arp 299-B-AT1, where \citet{mattila18} observed a high total radiated energy above $1.5\times10^{52}$ erg. The TDE happened in a nucleus hosting a known AGN and the transient showed a very significant, slowly evolving IR emission coming from the dust surrounding the AGN.

\subsection{Tidal disruption of a Neutron Star in the accretion disk}
Another possible scenario that could potentially explain the similarity between the flares observed in the three Gaia objects is that of a neutron star (NS) tidally disrupted by a stellar mass BH in the AGN accretion disk. The mass of the BH required for the tidal disruption depends on the NS equation of state and, to large extent, on the BH spin, see \citealt{lattimer76,shibata11}. While the electromagnetic signal coming from the NS-BH merger would be too weak to be visible against the high luminosity of the AGN, part of the NS material would be outflowing at relativistic speeds and hit the Hydrogen rich accretion disk, creating the observed flare. 
In this scenario, the flare is not directly related to the SMBH - as in the case of an accretion event - therefore it can more naturally explain the similar bolometric luminosity of the three events. A tidal encounter between a NS and a stellar mass BH in the proximity of an AGN accretion disk could be relatively common. Considering a distribution of NS and BH around the SMBH, a fraction of these objects will either have orbits on the disk plane, or will have orbits that intersect the disk and that will evolve to the orbital plane during the disk lifetime. Different objects within the disk will suffer different dynamical friction and will encounter each other at low relative velocities, favouring the creation of binaries \citep{stone17,mckernan12}.

\begin{figure}
    \includegraphics[width=1\columnwidth]{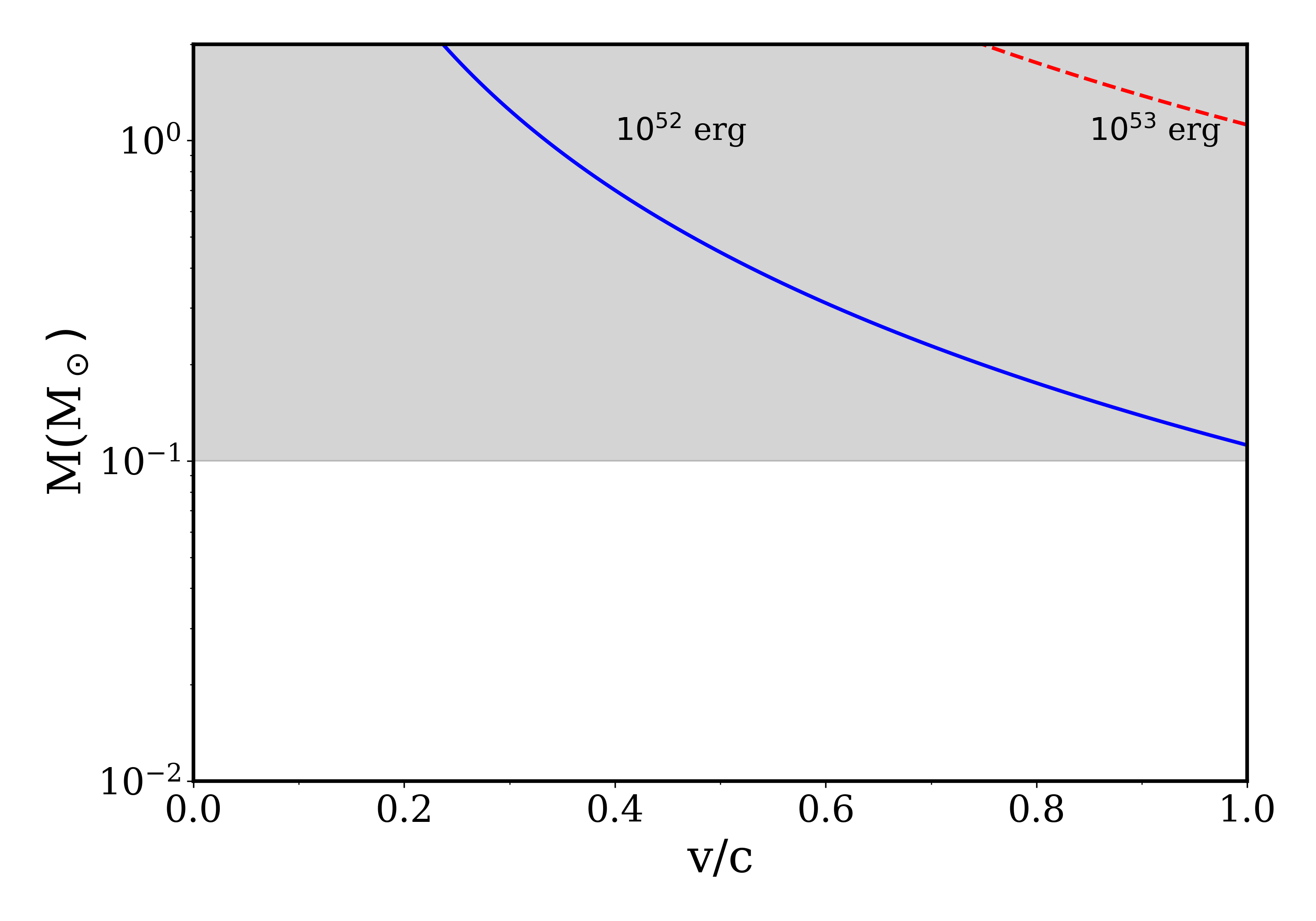}
    \caption{Energy ($1/2Mv^2$) as curves for $10^{52}$ erg (blue solid curve) and $10^{53}$ erg (red dashed curve) plotted in mass of the ejecta and its velocity. As shown in \citet{barbieri19}, we don't expect more than $0.1 M_\odot$ to be available for the NS ejecta, this limit is represented by the gray shaded area in the plot.}
    \label{fig:ns-tde}
\end{figure}

In Fig. \ref{fig:ns-tde} we show a simple calculation of the energy released in such a scenario. On the Y-axis we have the mass in the outflow (a fraction of the total NS mass) and on the X-axis the velocity of the outflow. The solid lines represent the kinetic energy ($1/2Mv^2$) of the outflowing material (with $M$ its mass and $v$ its velocity).

In this scenario, to reach an energy release of the order of $10^{52}$ erg, we would need a fraction of the NS material, around 10\%, to be traveling at $\sim$0.6 c, taking into account relativistic boosting, which is relevant at the considered velocity, that would help reach the required energy output. According to the calculations performed in \citet{barbieri19}, 0.1$M_\odot$ and $0.6c$ are the maximum values for the ejecta mass and velocity, respectively.

Gaia16aax released a total amount of energy of $\sim3\times10^{52}$ erg, therefore the energy provided by the dynamical ejecta is not enough to explain the emitted energy. \citealt{barbieri19} suggests that a jet could also carry O($10^{52}{\rm erg}$) and in \citet{deaton13}, a short-lived neutron-rich accretion disk around a rapidly spinning stellar mass black hole can have thermal energy also of O($10^{52}{\rm erg}$). Thus, we regard $\sim 3 \times 10^{52} {\rm erg}$ as a reasonable upper limit for the energy output from a NS TDE, requiring a jet, fast and relatively massive and high-velocity ejecta and a rapidly spinning BH.

As said, the appeal of this picture lies in that it naturally explains the similar bolometric luminosity of the three events, something which is more difficult to do if it is related to accretion onto the central BH. The energy released during the Gaia16aax outburst (and, by extension, of the other Gaia objects), though, stretches this scenario, since its energy release is consistent with the upper limit calculated for this scenario. However, the total energy output of Gaia16aax could be underestimated, as our calculation of the bolometric luminosity was done mainly using the optical photometry from Gaia. If this is the case, the scenario discussed in this section must be ruled out.

\section{Summary}
We present multiwavelength follow up of Gaia16aax, a nuclear transient discovered by the Gaia science alerts program. The transient is coincident with a quasar at $z=0.25$ (SDSS J143418.47+491236). The target brightened by more than 1 magnitude in the optical over less than a year and went back to its pre-outburst level over more than two years. Variability of similar amplitude has been detected also in the NIR (with a time delay of $\sim$140 days), UV and X-rays. The most striking property of this object is the shape of the \ha~ and \hb~ broad emission lines. These lines were present in the SDSS pre-outburst spectrum, but during the outburst they show two distinct peaks, at different shifts with respect to the rest-frame wavelength. Gaia16aax is part of a group of three nuclear transients discovered by the Gaia science alerts program that showed nearly identical light curves. The other two are Gaia16aka and Gaia16ajq, for which a detailed follow-up is not available. The three objects show similarities also in their spectra.

We discuss various scenarios present in the literature to explain large amplitude flares in AGNs to try to explain the observed properties of the outburst of Gaia16aax. The short timescale of the rise to peak and the total energy output are the most challenging properties to explain. Of these scenarios, some are easily ruled out: a microlensing event, a SN explosion, and variable absorption in the line of sight cannot explain the observed properties. Other phenomena are more promising, although none can explain all the observed properties in a straightforward manner. The outburst of Gaia16aax can be explained by a change in the accretion flow onto the central BH. The low Eddington ratio of Gaia16aax is a characteristic shared by many other quasars that showed variability of similar amplitude. But, in the theoretical framework of accretion disk physics, it is difficult to explain the rapid rise shown by Gaia16aax, as the time scales that govern the dynamics of an accretion disk are much longer than those at play in Gaia16aax. We review some proposed mechanisms in the literature for variability in the inner part of the accretion disk, finding that, with the aid of some magnetic wind-driven loss of angular momentum, a high amplitude variability on timescales of $\sim$1 year is possible, albeit the frequency of these episodes and their dependency on the disk properties remain ucertain.

The outburst could have been caused by a tidal disruption event. Given the high mass of the BH, to have a tidal disruption event the disrupted star needs to have mass $\geq1M_\odot$. The presence of star formation may allow for this. Certainly for the stars of around a solar mass the BH must be rapidly spinning. This would help explain both the high energy release of the event and the short timescale. The interaction between the debris stream and the accretion disk could help explain the shape of the light curve and the shape of the emission lines. The encounter between the debris stream and the accretion disk should give rise to different observable properties in the three Gaia objects, given the different SMBH masses and the dependence of such encounters on multiple parameters (e.g.~density of the stream, incidence angle). The similar timescale and peak brightness of the three transients question the validity of this scenario. We cannot exclude that the detection of the three similarly shaped flares is a product of the Gaia selection criteria and that we are missing more TDEs in AGNs. It is important to note that transient candidates detected by Gaia are vetted by eye before being published \citep{kostrzewa18} and it is therefore very difficult to gauge the selection function of Gaia.

We also explore the possibility that Gaia16aax is due to a NS -- BH merger happening in the AGN disk. This scenario would help explain the three different transients independently of the central SMBH properties, but the high total energy output of Gaia16aax is on the high side of what is conceivably produced in this picture.

\section*{Acknowledgements}
GC, PGJ and ZKR acknowledge support from European Research Council Consolidator Grant 647208. 
MF is supported by a Royal Society - Science Foundation Ireland University Research Fellowship. FO acknowledges the support of the H2020 European Hemera program, grant agreement No 730970. JH acknowledges financial support from the Finnish Cultural Foundation and the Vilho, Yrj{\"o} and Kalle V{\"a}is{\"a}l{\"a} Foundation (of the Finnish Academy of Science and Letters). BM \& KESF are supported by NSF grant 1831412. CJN is supported by the Science and Technology Facilities Council (STFC) (grant number ST/M005917/1). We thank the referee for their useful comments and suggestions. We acknowledge ESA Gaia, DPAC and the Photometric Science Alerts Team (http://gsaweb.ast.cam.ac.uk/alerts). NUTS is supported in part by IDA (The Instrument Centre for Danish Astronomy). The data presented here were obtained [in part] with ALFOSC, which is provided by the Instituto de Astrofisica de Andalucia (IAA) under a joint agreement with the University of Copenhagen and NOTSA. Based [in part] on observations made with the Nordic Optical Telescope, operated by the Nordic Optical Telescope Scientific Association at the Observatorio del Roque de los Muchachos, La Palma, Spain, of the Instituto de Astrofisica de Canarias. The Liverpool Telescope is operated on the island of La Palma by Liverpool John Moores University in the Spanish Observatorio del Roque de los Muchachos of the Instituto de Astrofisica de Canarias with financial support from the UK Science and Technology Facilities Council. The CSS survey is funded by the National Aeronautics and Space Administration under Grant No. NNG05GF22G issued through the Science Mission Directorate Near-Earth Objects Observations Program. The CRTS survey is supported by the U.S.~National Science Foundation under grants AST-0909182 and AST-1313422. Based on observations obtained with XMM-Newton, an ESA science mission with instruments and contributions directly funded by ESA Member States and NASA.




\bibliographystyle{mnras}
\bibliography{gaia16aax.bib} 




\appendix
\section{Additional material}

\begin{figure*}
\centering
\subfloat[20160209]{\label{spec_hb_a}\includegraphics[width=.33\textwidth]{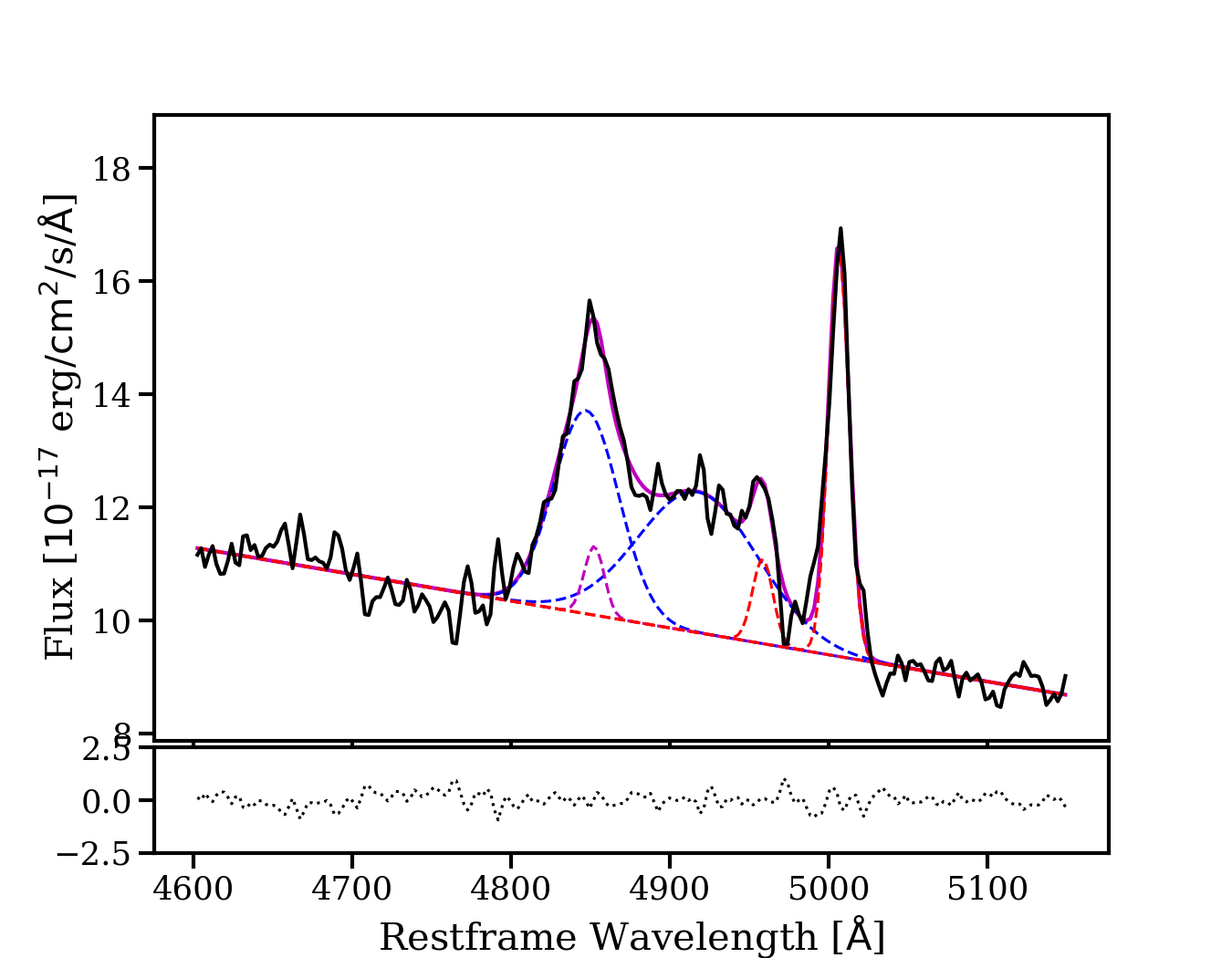}}
\subfloat[20160701]{\label{spec_hb_b}\includegraphics[width=.33\textwidth]{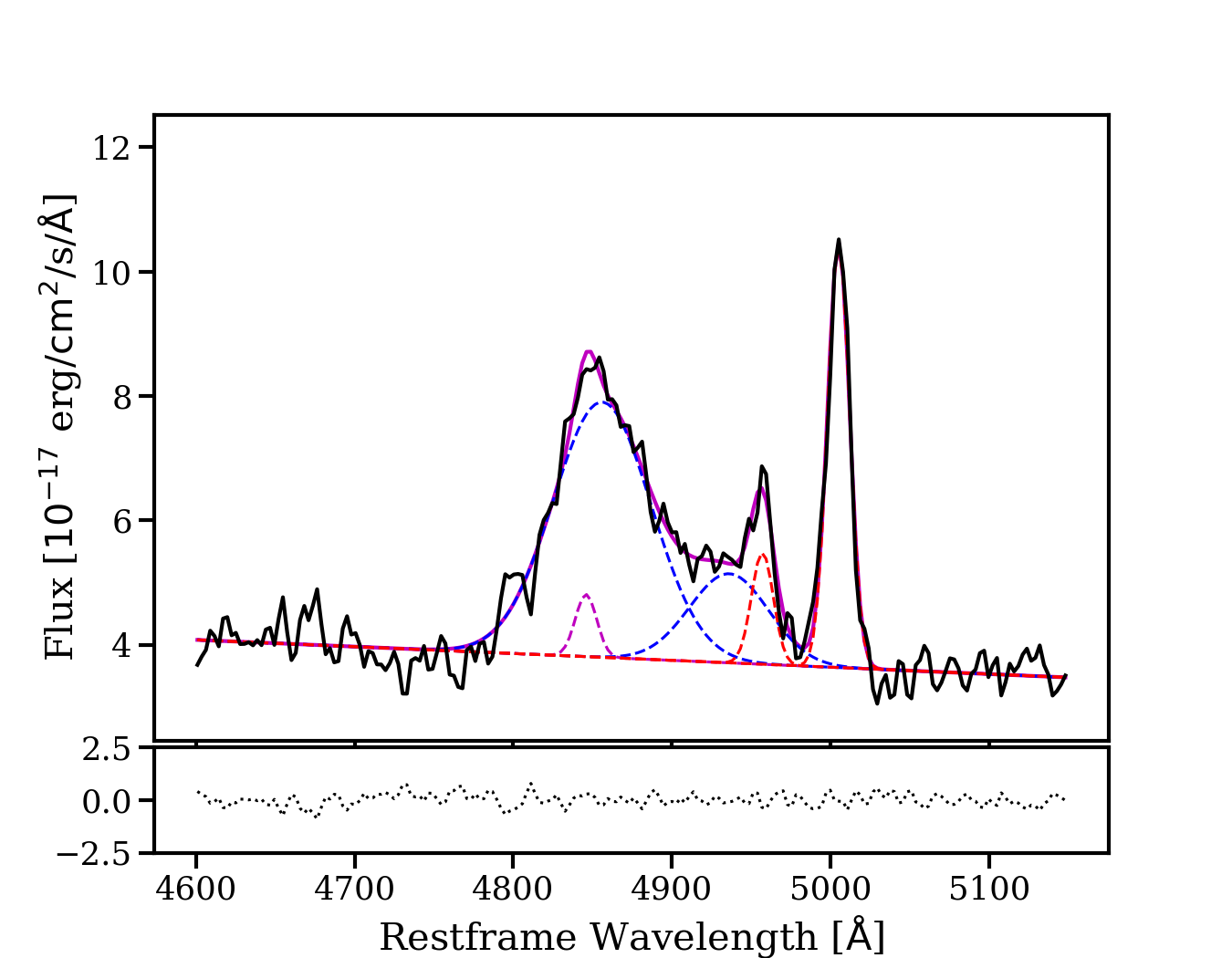}}
\subfloat[20160731]{\label{spec_hb_c}\includegraphics[width=.33\textwidth]{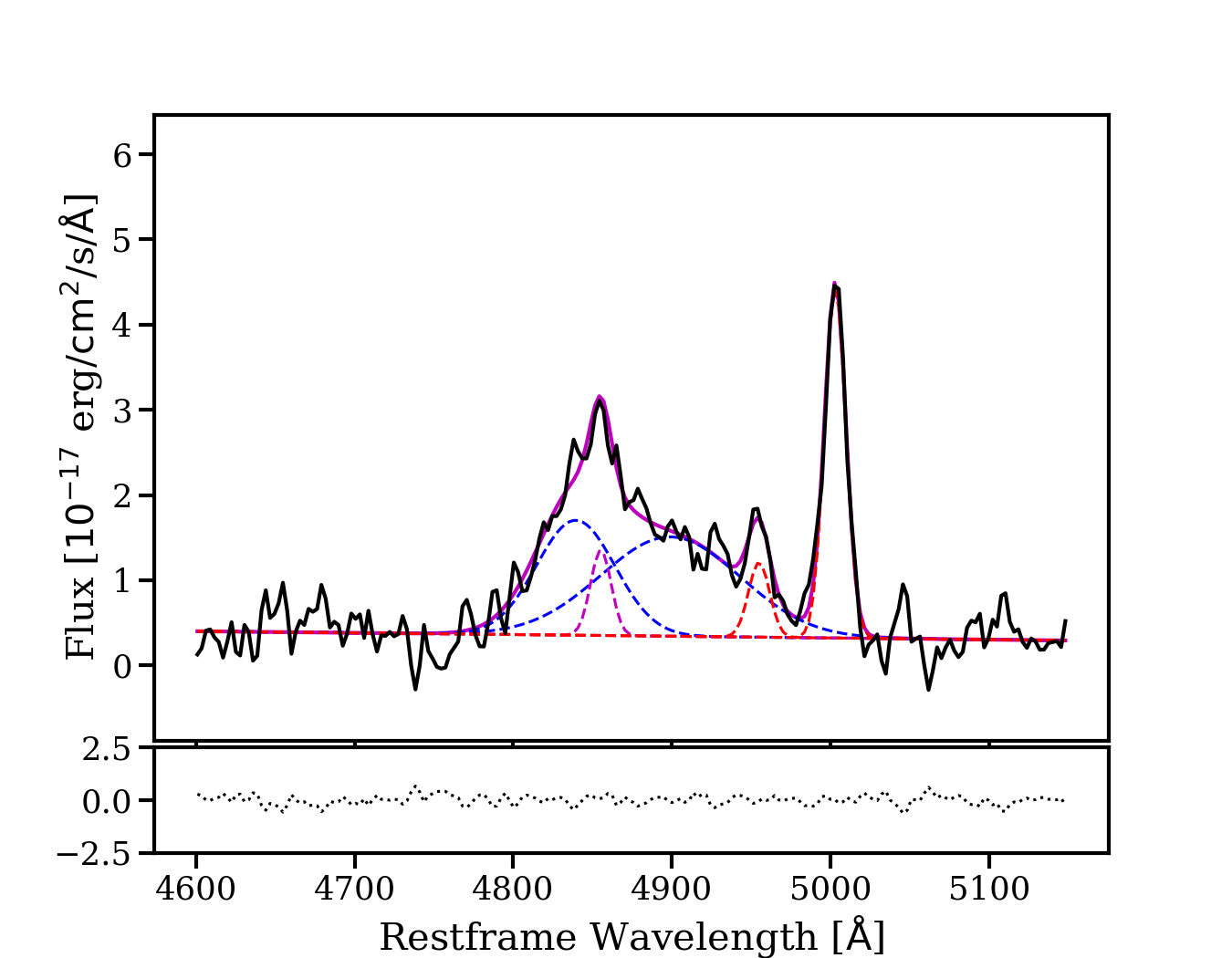}}\\
\subfloat[20161209]{\label{spec_hb_d}\includegraphics[width=.33\textwidth]{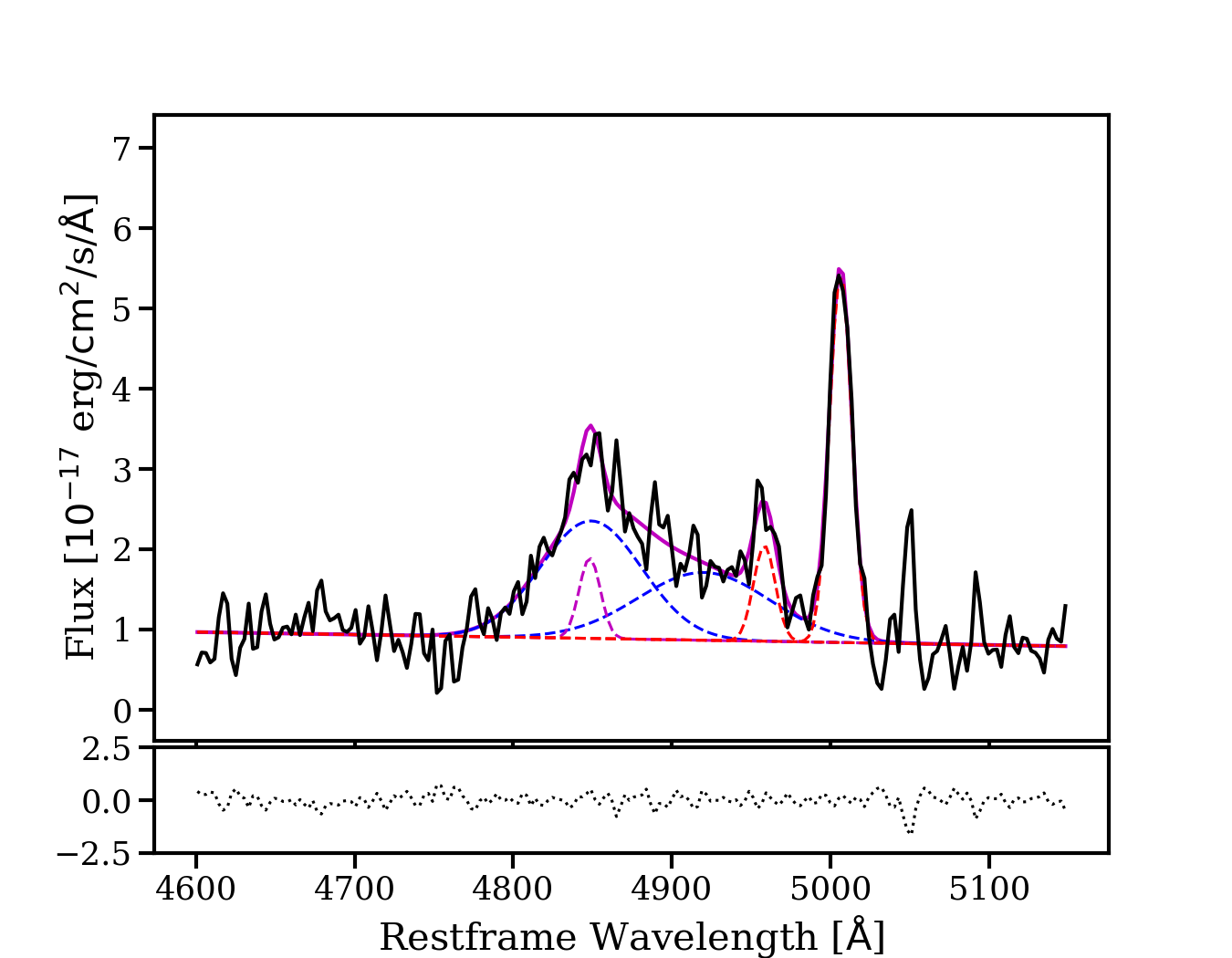}}
\subfloat[20170401]{\label{spec_hb_f}\includegraphics[width=.33\textwidth]{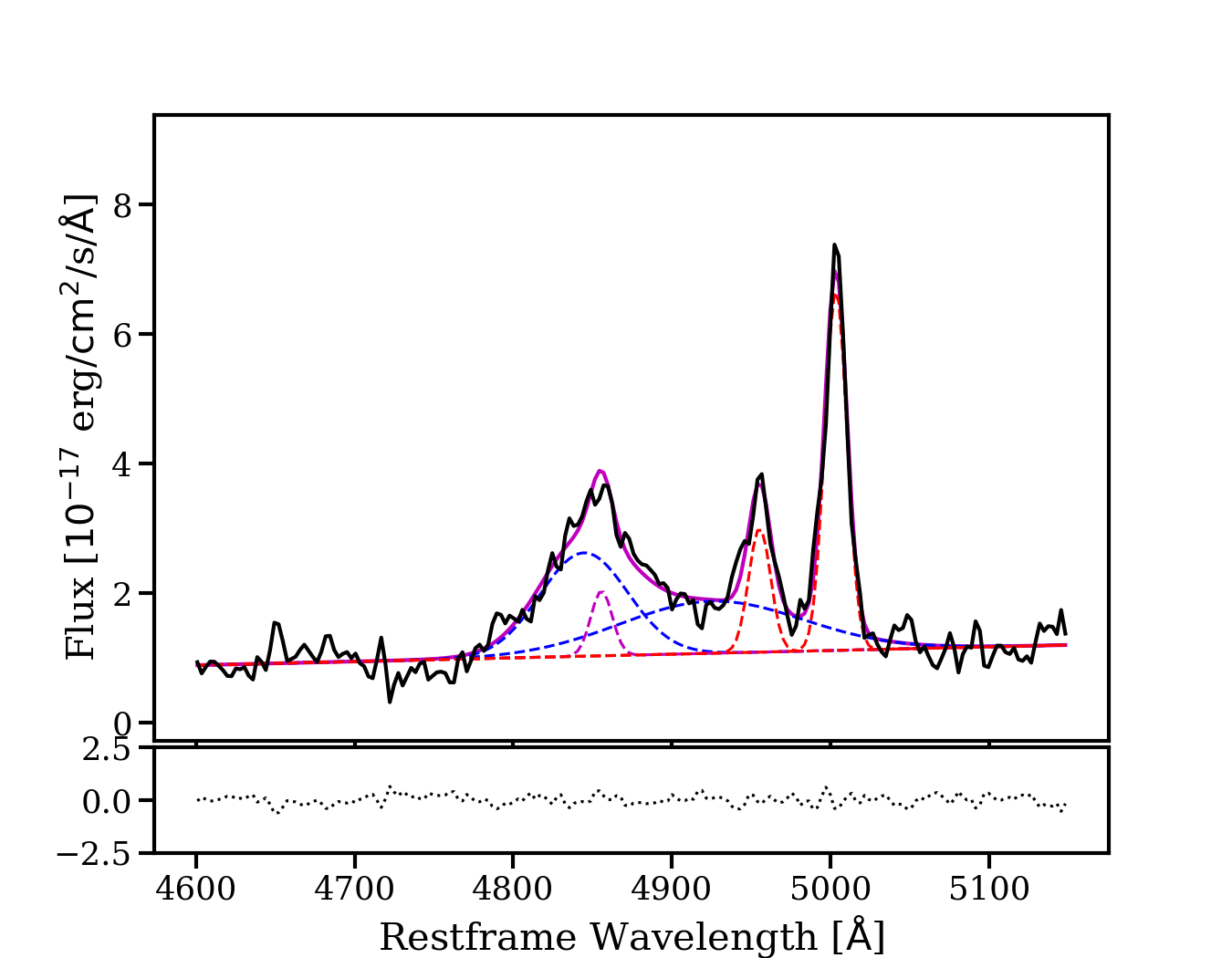}}
\subfloat[20170701]{\label{spec_hb_g}\includegraphics[width=.33\textwidth]{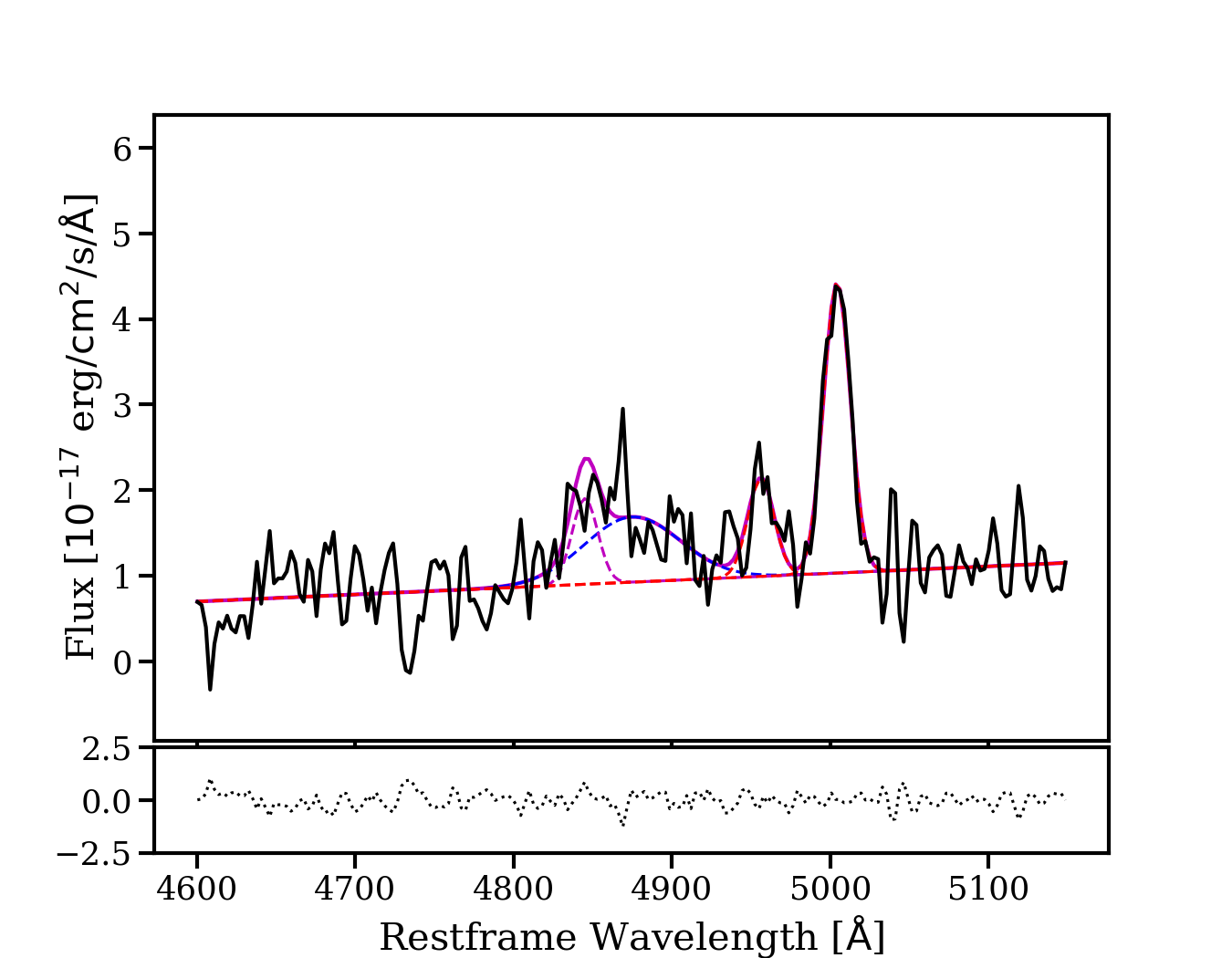}}\\
\subfloat[20180719]{\label{spec_hb_h}\includegraphics[width=.33\textwidth]{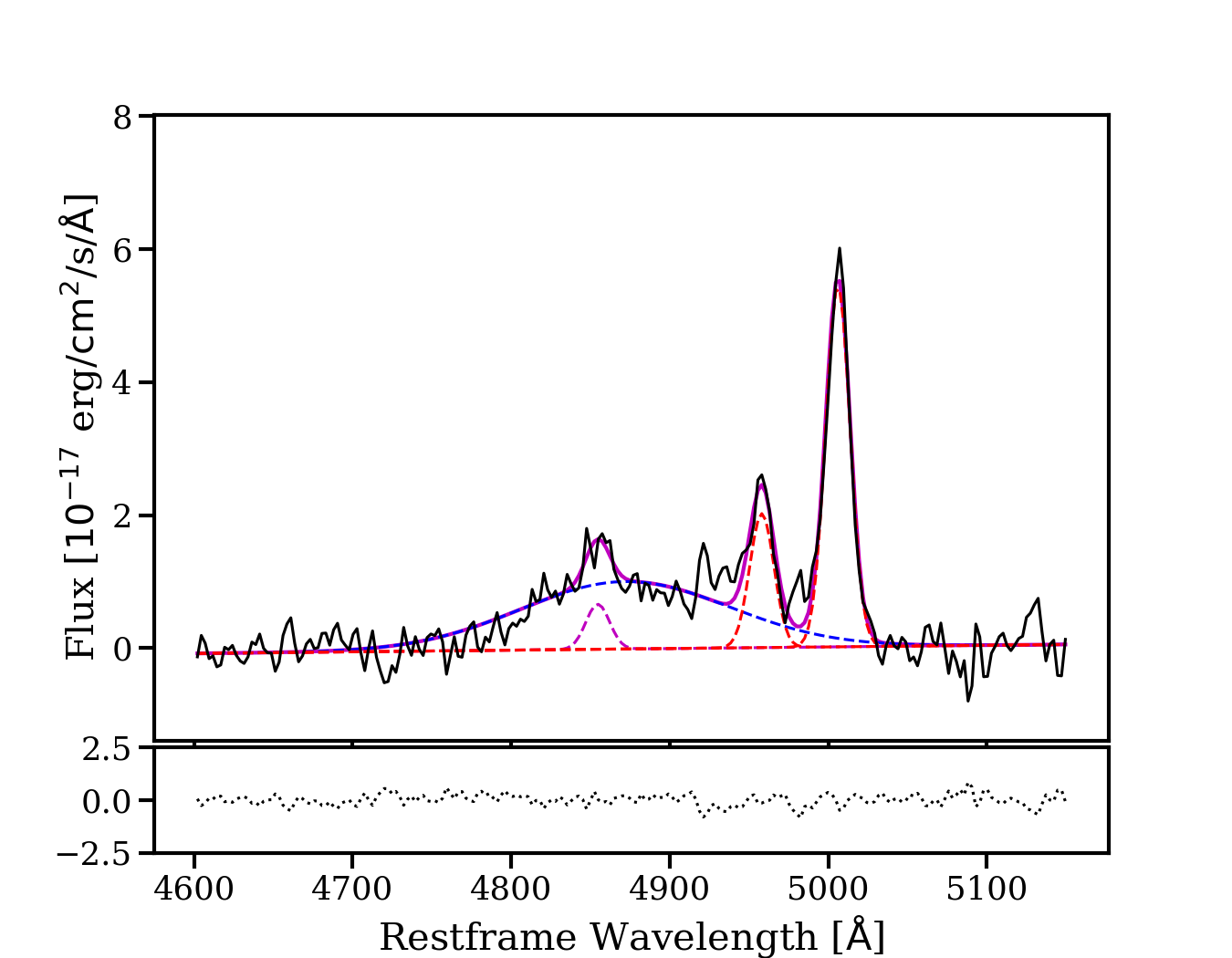}}
\subfloat[SDSS - 20020705]{\label{spec_hb_i}\includegraphics[width=.33\textwidth]{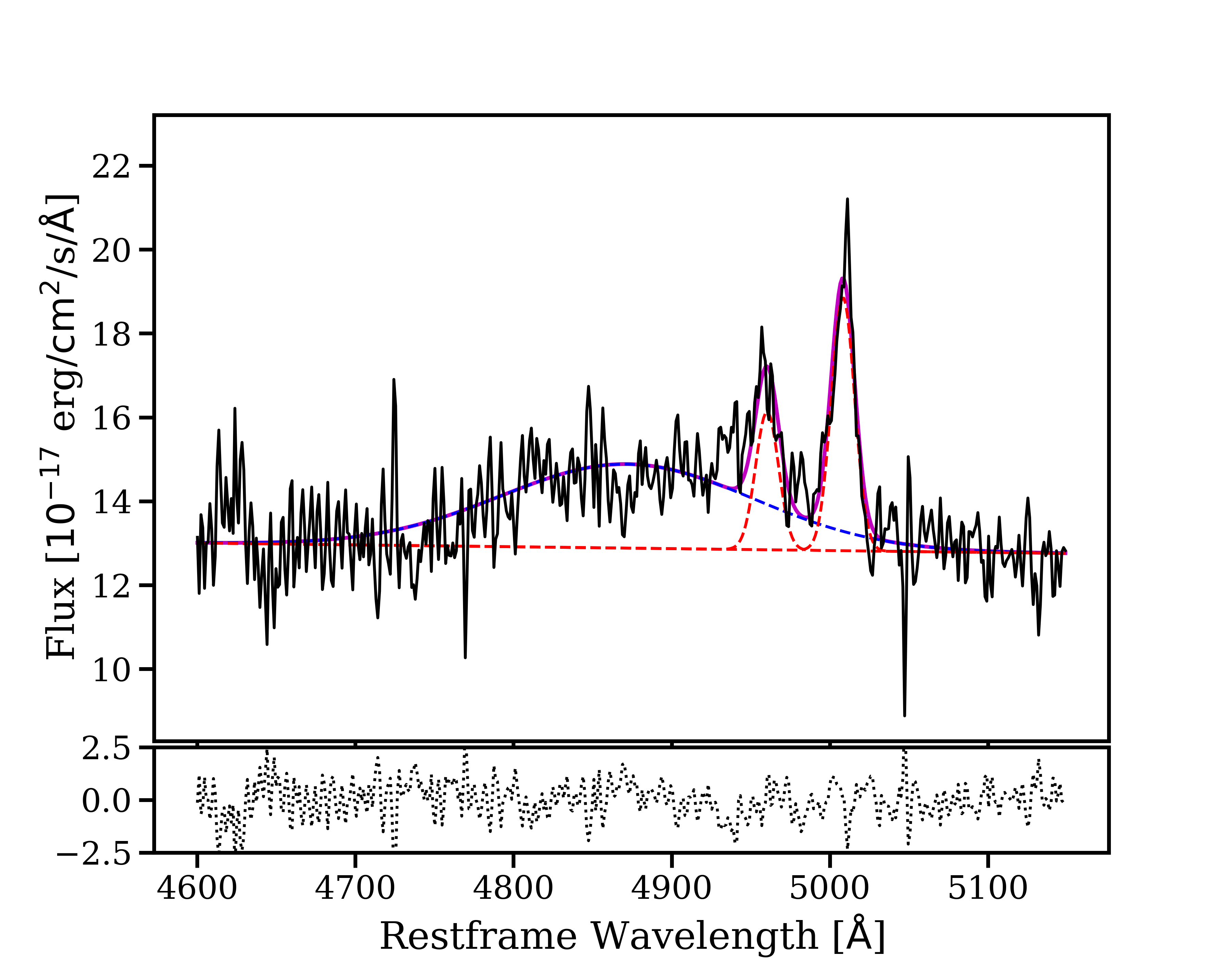}}
\caption{Evolution in time of the \hb~line region with the fit to the various components with multiple Gaussian curves. In all the panels the blue dashed lines are the two \hb~components, the magenta dashed line is the narrow \hb~component and the red dashed lines are the two \oiii emission lines. The spectra of 2017 January 26 is not shown since a cosmic ray hit the CCD on top of the \hb~line, making the results of the fit unreliable even after several attempts at correcting the data. Panel g is the \hb~line region at an epoch when the Gaia lightcurve has reached the pre-outburst level. The double-peaked shape of \hb~shown in outburst is not clearly present anymore at the last two epochs (panels f and g). The last panel is the SDSS spectrum, obtained well before the outburst start.}
\label{fig:hb_mosaic}
\end{figure*}

\begin{figure*}
\centering
\subfloat[20160209]{\label{spec_ha_a}\includegraphics[width=.33\textwidth]{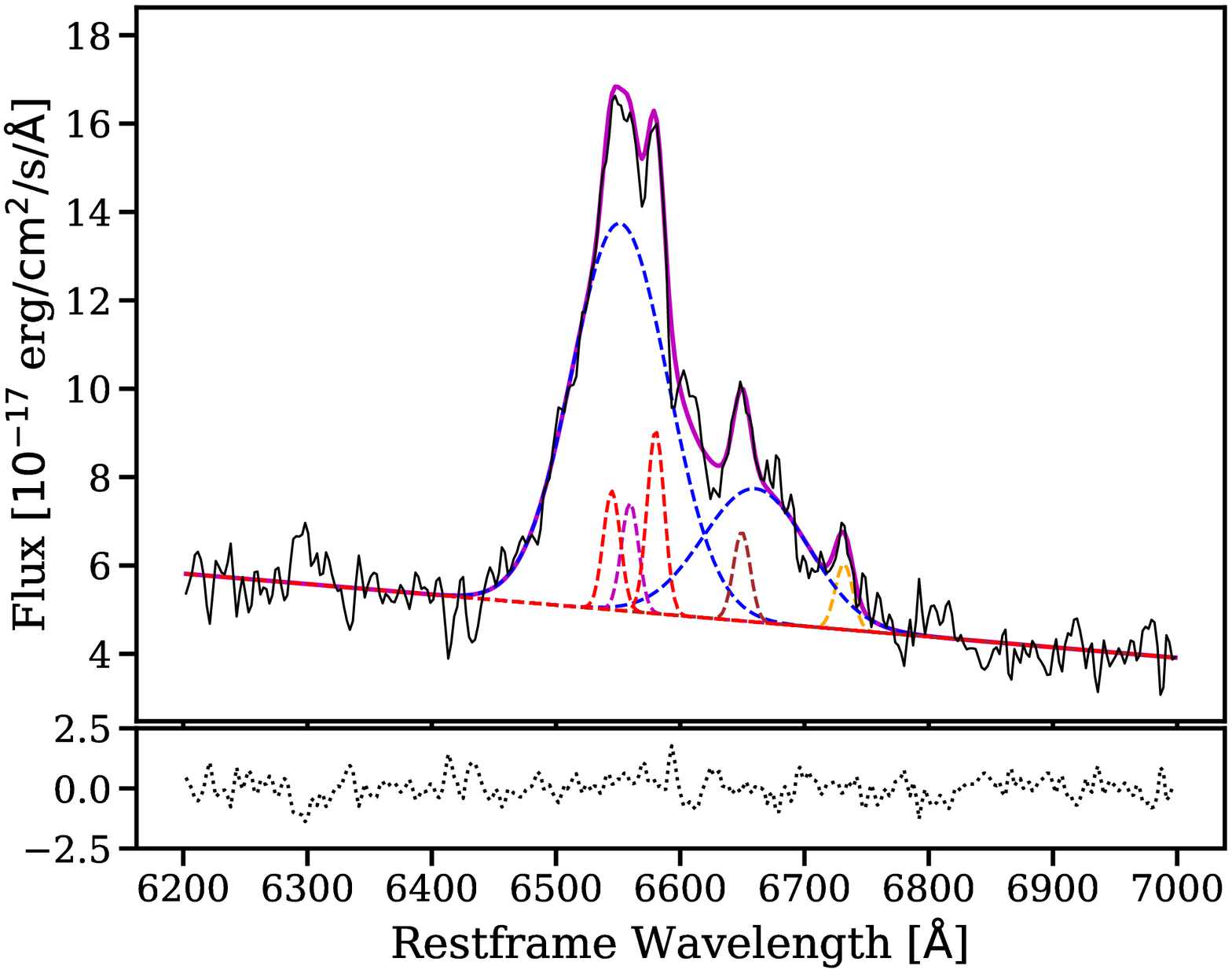}}\hfill
\subfloat[20160701]{\label{spec_ha_b}\includegraphics[width=.33\textwidth]{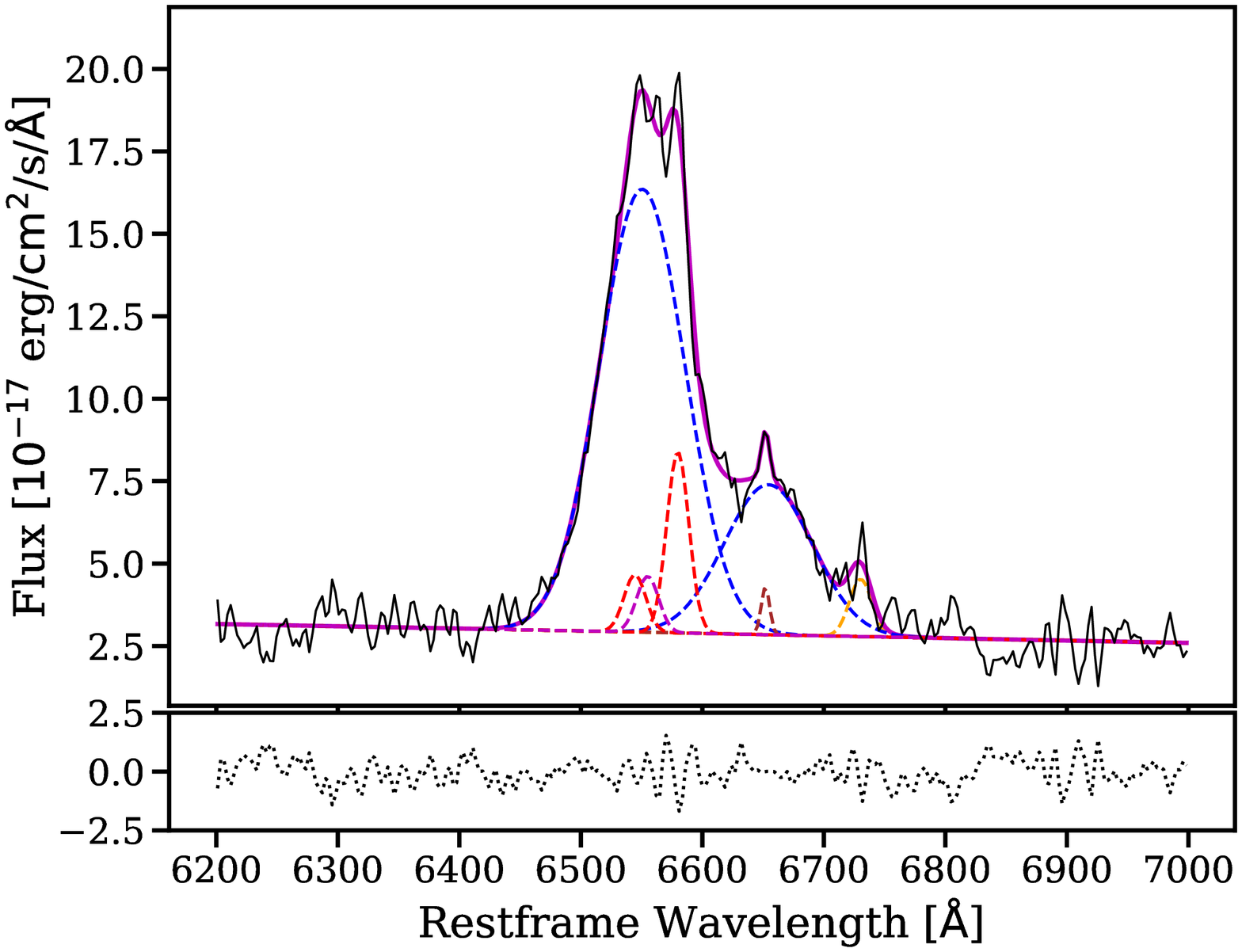}}\hfill
\subfloat[20160731]{\label{spec_ha_c}\includegraphics[width=.33\textwidth]{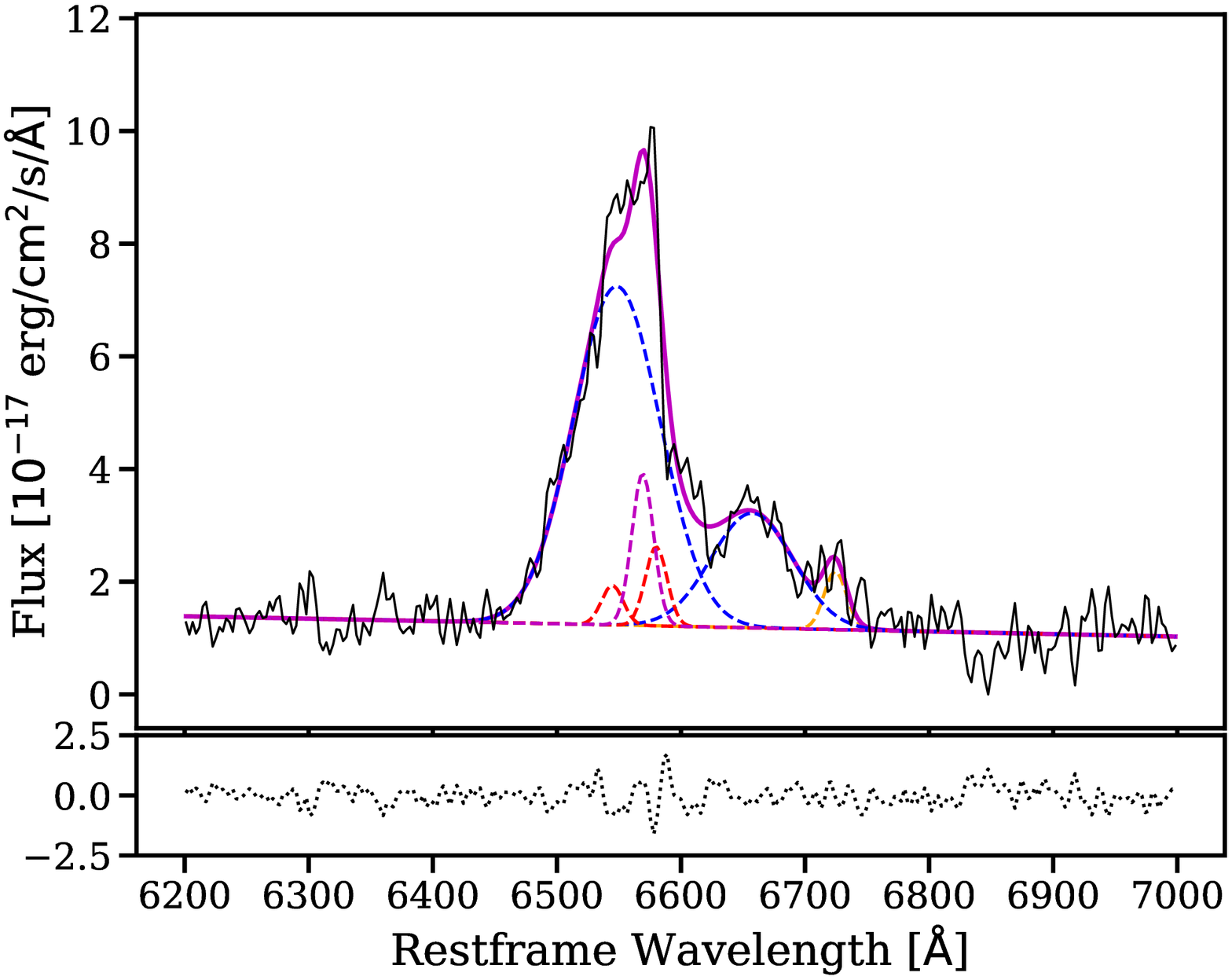}}\\
\subfloat[20161209]{\label{spec_ha_d}\includegraphics[width=.33\textwidth]{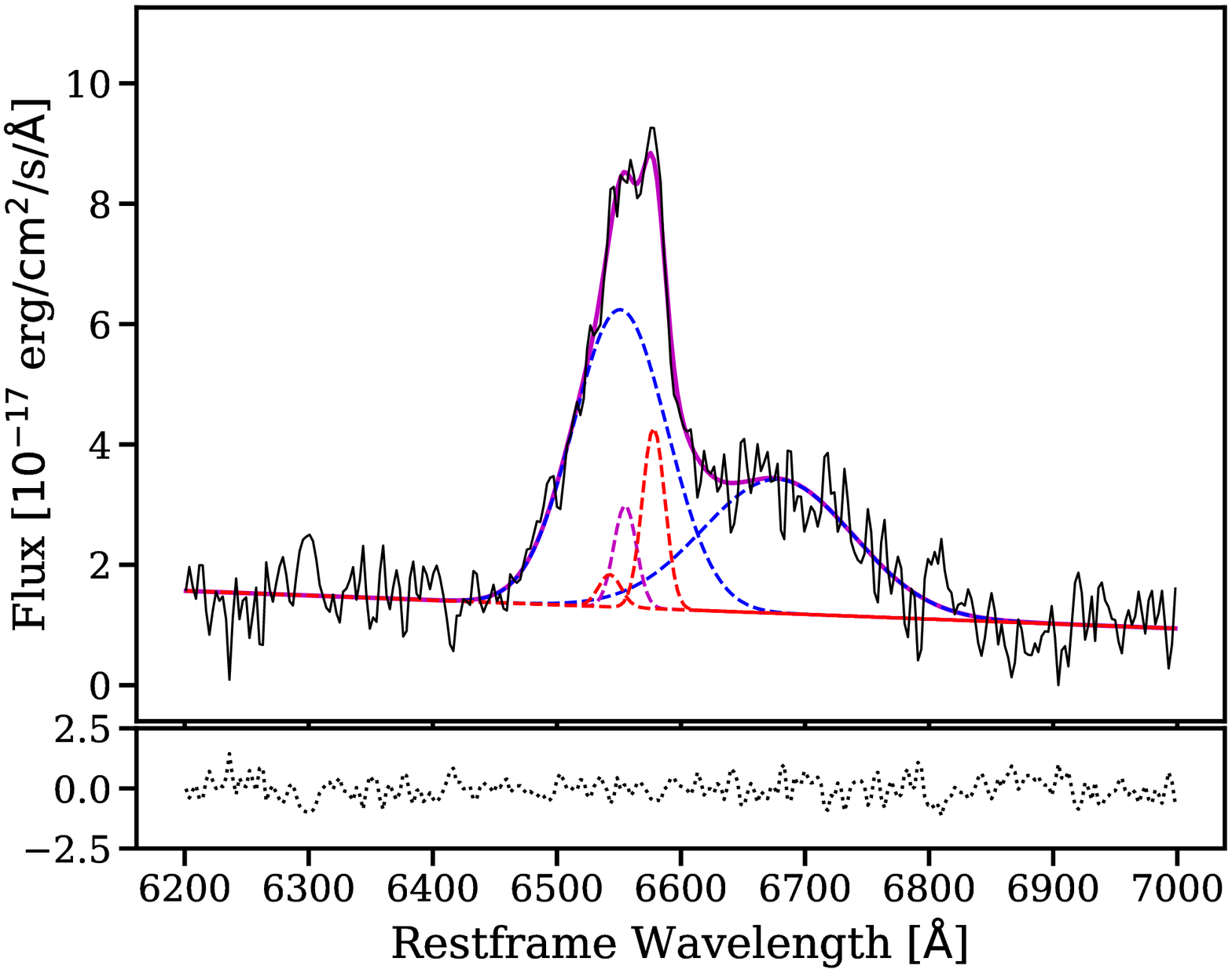}}\hfill
\subfloat[20170126]{\label{spec_ha_e}\includegraphics[width=.33\textwidth]{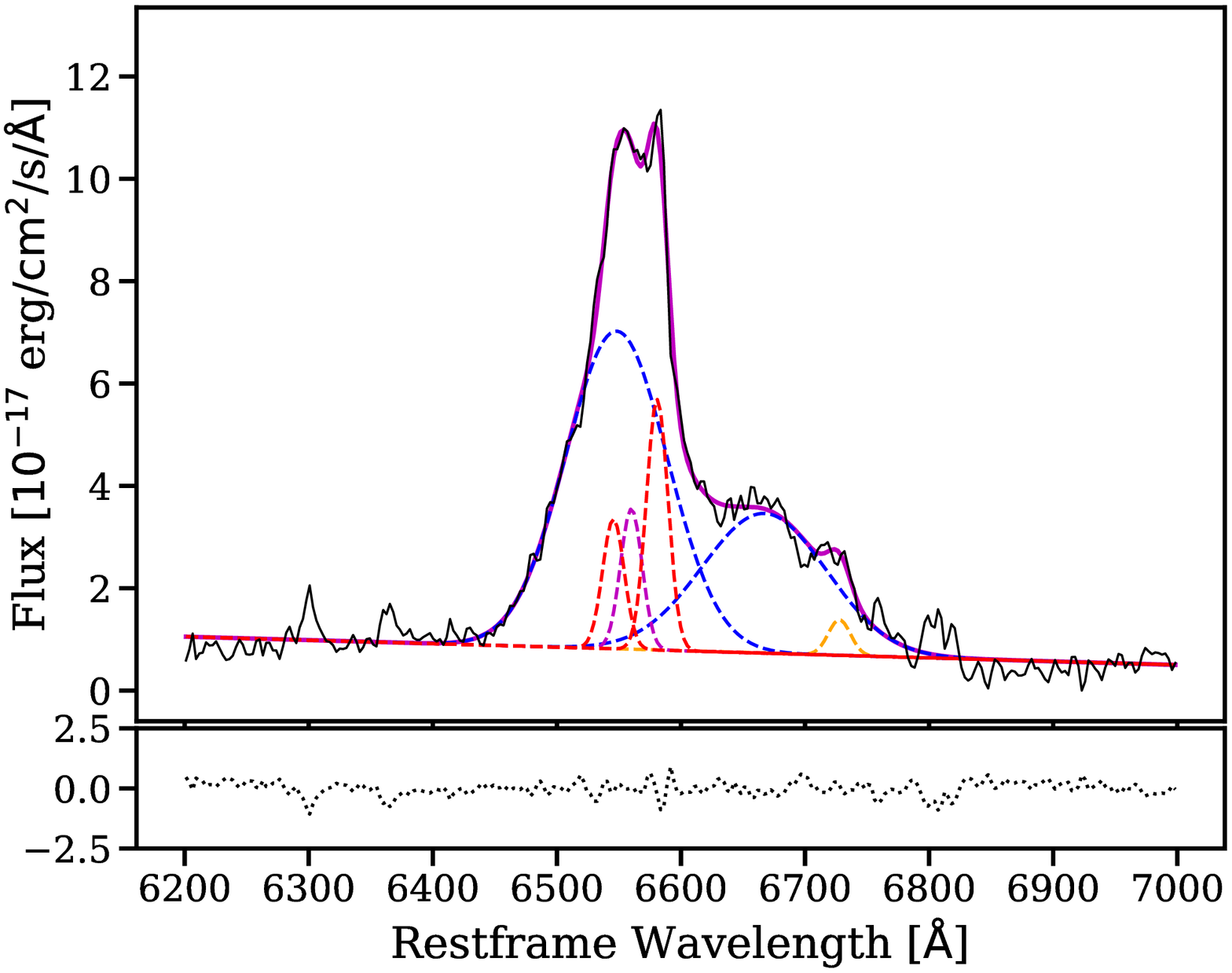}}\hfill
\subfloat[20170401]{\label{spec_ha_f}\includegraphics[width=.33\textwidth]{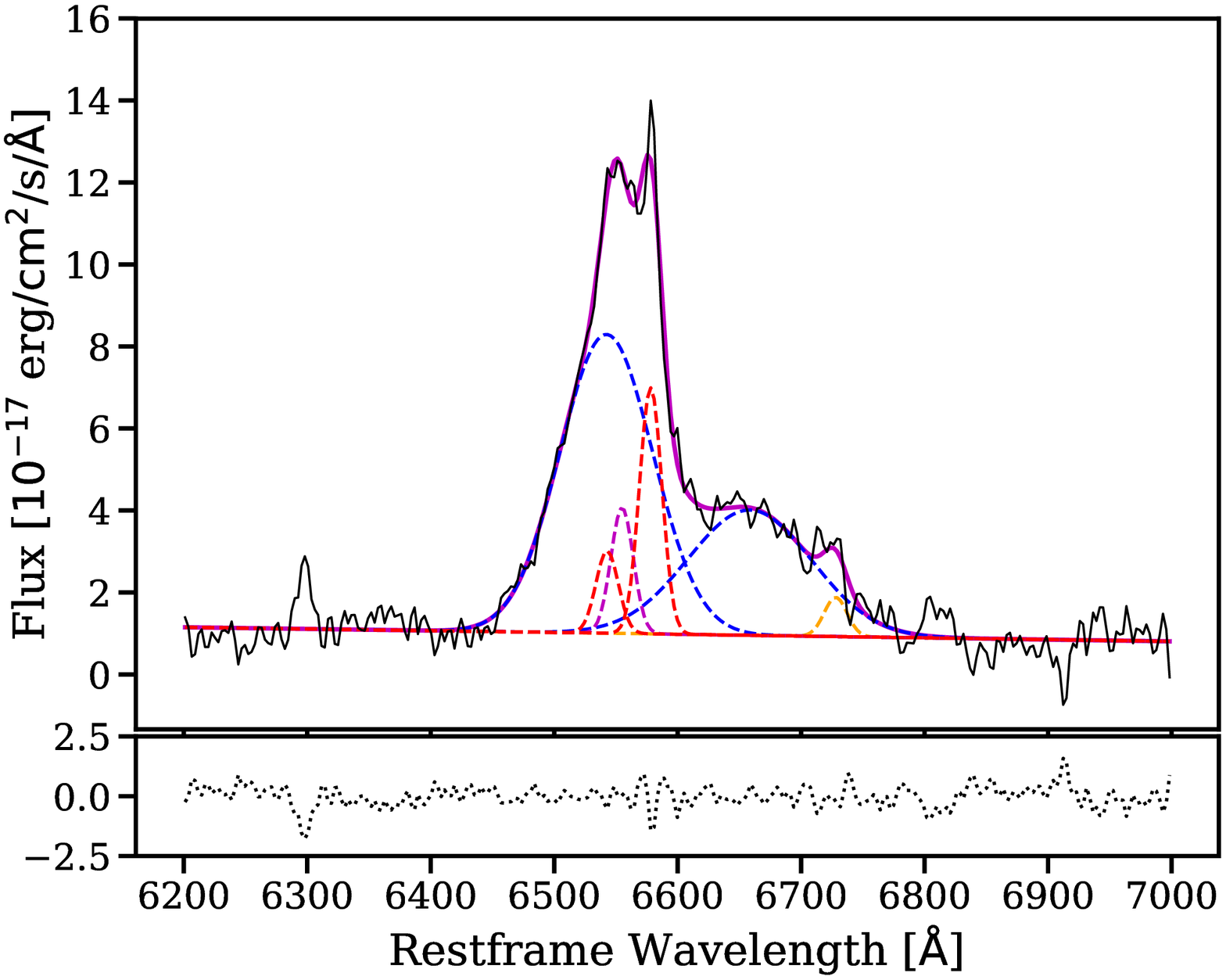}}\\
\subfloat[20170701]{\label{spec_ha_g}\includegraphics[width=.33\textwidth]{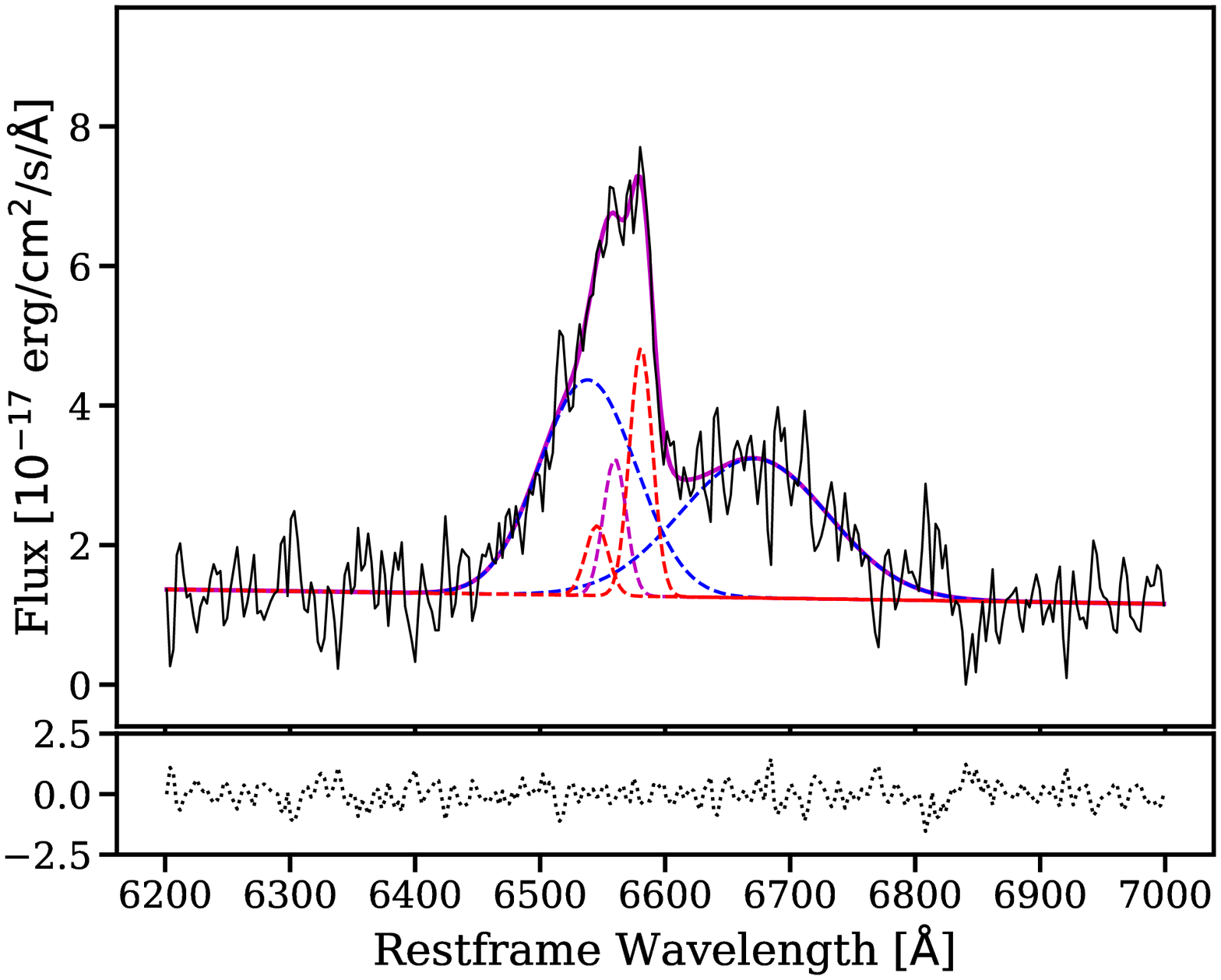}}
\subfloat[20180719]{\label{spec_ha_h}\includegraphics[width=.33\textwidth]{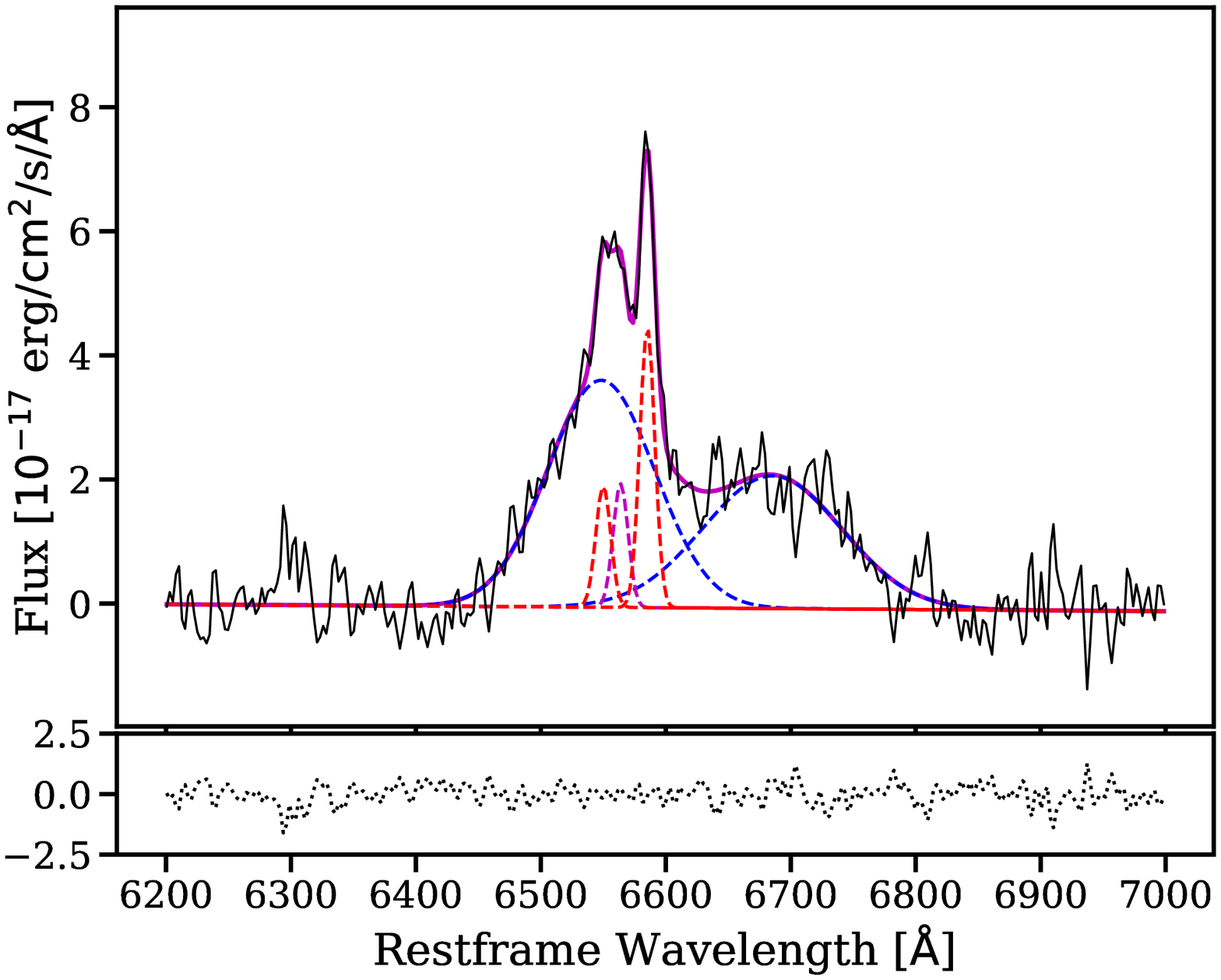}}
\subfloat[SDSS - 20020705]{\label{spec_ha_l}\includegraphics[width=.33\textwidth]{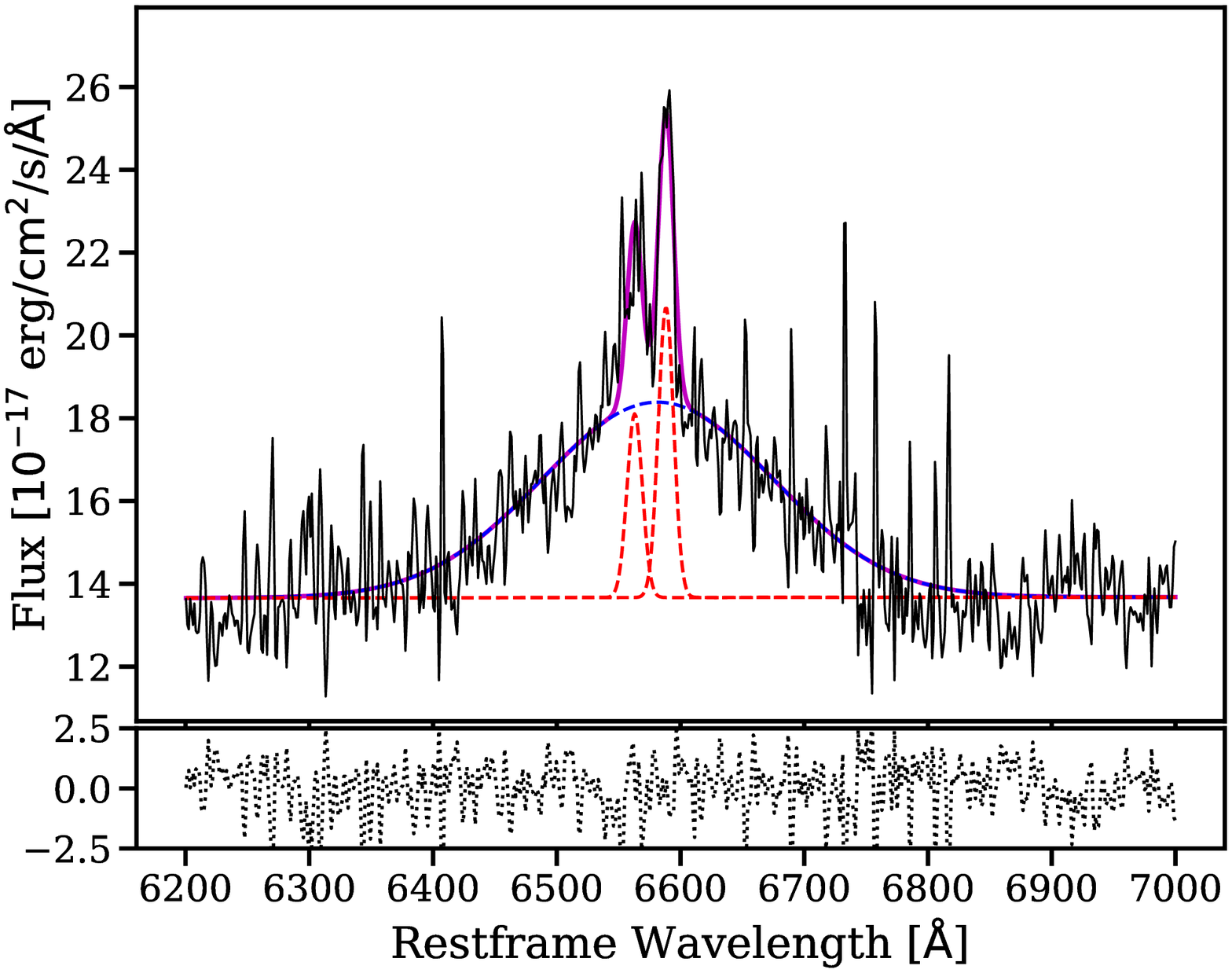}}
\caption{Evolution in time of the \ha~line region with the fit to the various components with multiple Gaussian curves. In all panels the red dashed lines are the two \nii lines, the blue dashed lines are the two \ha~components, the magenta dashed line is the \ha~ narrow component, the brown dashed line is the \ion{He}{i} line (only present in the first two panels, a and b) and the orange dashed line is the blend of the two [\ion{S}{ii}] lines. Panel h is the spectrum taken after the object returned to its pre-outburst state, according to the Gaia lightcurve.}
\label{fig:ha_mosaic}
\end{figure*}


\begin{landscape}
\begin{table}
\caption{Magnitude values in the Optical and NIR.}
\label{tab:photometry}
\begin{center}
\begin{tabular}{lccccccccccc}\hline
MJD           &  \textit{u}  &  B     & V & \textit{g} & \textit{r} & \textit{i}  & \textit{z} & \textit{y}  & J & H  & $\rm K_s$\\
$\rm [d]$ (1) & (2)          & (3)            &(4)             & (5)            & (6)            & (7)        & (8) & (9) & (10) & (11) & (12)\\
\hline
51\,671.30 (13)& $\cdots$ & $\cdots$ & $\cdots$ & $\cdots$ & $\cdots$ & $\cdots$ & $\cdots$ &$\cdots$ & 15.96$\pm$0.09 & 15.47$\pm$0.14 & 14.64$\pm$0.10 \\
52\,402.00 (14) & 19.75$\pm$0.04 &  $\cdots$& $\cdots$ & 19.47$\pm$0.02 & 18.65$\pm$0.02 & 18.13$\pm$0.01 & 17.71$\pm$0.02 &$\cdots$&$\cdots$&$\cdots$&$\cdots$\\ 
56\,938.05 (15)   & $\cdots$ & $\cdots$ & $\cdots$ & 19.69$\pm$0.02 & 18.88$\pm$0.02 & 18.34$\pm$0.05 & 17.99$\pm$0.06 & 17.76$\pm$0.05&$\cdots$&$\cdots$&$\cdots$\\
\hline
57\,462.24 & 18.94$\pm$0.09 & 18.95$\pm$0.09 & 19.01$\pm$0.06 & 19.17$\pm$0.08 & 19.25$\pm$0.06 & 18.97$\pm$0.05 & $\cdots$       &$\cdots$&$\cdots$&$\cdots$&$\cdots$\\
57\,476.24 & 19.12$\pm$0.07 & 19.07$\pm$0.06 & 19.05$\pm$0.06 & 19.25$\pm$0.07 & 19.30$\pm$0.04 & 19.08$\pm$0.04 & 18.70$\pm$0.04 &$\cdots$&$\cdots$&$\cdots$&$\cdots$\\
57\,491.01 & 19.01$\pm$0.06 & 19.15$\pm$0.06 & 19.18$\pm$0.07 & 19.40$\pm$0.08 & 19.35$\pm$0.05 & 19.90$\pm$0.04 & 19.23$\pm$0.08 &$\cdots$&$\cdots$&$\cdots$&$\cdots$ \\
57\,505.03 & 19.21$\pm$0.07 & 19.22$\pm$0.07 & 19.31$\pm$0.09 & 19.54$\pm$0.08 & 19.42$\pm$0.07 & 19.12$\pm$0.07 & 18.91$\pm$0.11 &$\cdots$&$\cdots$&$\cdots$&$\cdots$ \\
57\,509.04 &     $\cdots$   & $\cdots$ & $\cdots$ & $\cdots$ & $\cdots$ & $\cdots$ & $\cdots$ & $\cdots$ &  17.92$\pm$0.04 & 16.79$\pm$0.12 & 15.69$\pm$0.06 \\
57\,530.02 &     $\cdots$   & $\cdots$ & $\cdots$ & $\cdots$ & $\cdots$ & $\cdots$ & $\cdots$ & $\cdots$ &  17.81$\pm$0.03 & 16.61$\pm$0.08 & 15.63$\pm$0.06 \\
57\,551.01 &     $\cdots$   & $\cdots$ & $\cdots$ & $\cdots$ & $\cdots$ & $\cdots$ & $\cdots$ & $\cdots$ &  17.85$\pm$0.04 & 16.62$\pm$0.12 & 15.45$\pm$0.06 \\
57\,560.91 & 19.53$\pm$0.08 & 19.64$\pm$0.06 & 19.74$\pm$0.07 & 19.90$\pm$0.10 & 19.84$\pm$0.05 & 19.38$\pm$0.05 & 19.33$\pm$0.12&$\cdots$&$\cdots$&$\cdots$&$\cdots$ \\
57\,588.95 & 19.32$\pm$0.39 & 19.72$\pm$0.12 & 19.99$\pm$0.10 & 20.00$\pm$0.30 & 20.05$\pm$0.07 & 19.42$\pm$0.10 & 19.87$\pm$0.21&$\cdots$&$\cdots$&$\cdots$&$\cdots$ \\
57\,611.95 &     $\cdots$   & $\cdots$ & $\cdots$ & $\cdots$ & $\cdots$ & $\cdots$ & $\cdots$ & $\cdots$ &  17.97$\pm$0.08 & 16.65$\pm$0.23 & 15.35$\pm$0.05 \\
57\,738.25 &     $\cdots$   & $\cdots$ & $\cdots$ & $\cdots$ & $\cdots$ & $\cdots$ & $\cdots$ & $\cdots$ &  18.09$\pm$0.07 & 17.64$\pm$0.15 & 16.15$\pm$0.06 \\
57\,791.27 &     $\cdots$   & $\cdots$ & $\cdots$ & $\cdots$ & $\cdots$ & $\cdots$ & $\cdots$ & $\cdots$ &     $\cdots$    & 17.47$\pm$0.13 & 16.22$\pm$0.08 \\
57\,837.03 &     $\cdots$   & $\cdots$ & $\cdots$ & $\cdots$ & $\cdots$ & $\cdots$ & $\cdots$ & $\cdots$ &  19.18$\pm$0.12 & 17.91$\pm$0.29 & 16.58$\pm$0.11 \\
57\,937.97 &     $\cdots$   & $\cdots$ & $\cdots$ & $\cdots$ & $\cdots$ & $\cdots$ & $\cdots$ & $\cdots$ &  19.68$\pm$0.14 & 18.18$\pm$0.14 & 17.79$\pm$0.12 \\
\hline

\end{tabular}
\end{center}
 \textit{Notes:} (1) MJD date of observations; (2), (5), (6), (7), (8), (9) apparent magnitudes with 1-$\upsigma$ uncertainties in the optical bands \textit{u}, \textit{g}, \textit{r}, \textit{i}, \textit{z} and \textit{y}, in the AB system; (3) and (4) apparent magnitudes with 1-$\upsigma$ uncertainties in the optical bands B and V, in the Vega system; (10), (11) and (12) extinction-corrected apparent magnitudes with 1-$\upsigma$ uncertainties in the Near Infra-Red filters J, H and K$_s$, in the Vega system. The first three lines are the archival magnitudes from 2MASS (13), SDSS (14) and PanSTARRS (15), all the other values are the result of the image subtraction procedure. The symbol $\cdots$ indicates an epoch in which an observation for the specific filter was not available.
\end{table}
\begin{table}
\caption{Results from the line fitting procedure on \ha~ and \hb.}
\label{tab:spectroscopy}
\begin{center}
\begin{tabular}{lcccccccccccc}\hline
MJD$^{(1)}$ & \hba FWHM$^{(2)}$ & \hbb FWHM$^{(2)}$ & \hba flux$^{(3)}$ & \hbb flux$^{(3)}$ & \hba shift$^{(4)}$ & \hbb shift$^{(4)}$ & \haa FWHM$^{(2)}$ & \hab FWHM$^{(2)}$  &  \haa flux$^{(3)}$ & \hab flux$^{(3)}$ & \haa shift$^{(4)}$ & \hab shift$^{(4)}$\\
        			&    [\kms]      & [\kms] &[\unitflux]&[\unitflux]&[\kms] & [\kms] & [\kms] & [\kms] & [\unitflux] & [\unitflux] & [\kms] & [\kms]  \\
\hline 
57428.27         & 3134$\pm$206 & 4775$\pm$466  & 197$\pm$19 & 210$\pm$22 & -833$\pm$104  & 3593$\pm$201 &     4168$\pm$152 & 4464$\pm$541 & 850$\pm$51   & 318$\pm$46 & -520$\pm$114 & 4464$\pm$299 \\
57571.06         & 4340$\pm$238 & 3724$\pm$495  & 316$\pm$20 & 96$\pm$15  & -394$\pm$105  &  4492$\pm$292&     3778$\pm$120 & 3856$\pm$391 & 1180$\pm$70  & 414$\pm$47 & -560$\pm$88  & 4186$\pm$169 \\
57600.97         & 3801$\pm$482 & 5914$\pm$1696 & 112$\pm$34 & 107$\pm$35 & -1064$\pm$302 &  2905$\pm$1157&     3767$\pm$178 & 3482$\pm$493 & 526$\pm$42   & 168$\pm$26 & -667$\pm$117 & 4322$\pm$191 \\
57732.25         & 4590$\pm$445 & 6192$\pm$884  & 113$\pm$15 & 93$\pm$25  &  -795$\pm$133 & 3514$\pm$636     &     4069$\pm$212 & 6400$\pm$704 & 468$\pm$47   & 338$\pm$40 & -537$\pm$122 & 5265$\pm$336 \\
57780.26$^{(5)}$ &  $\cdots$    &   $\cdots$    &  $\cdots$  & $\cdots$   &  $\cdots$     & $\cdots$ &     4410$\pm$141 & 5294$\pm$161 & 637$\pm$34   & 342$\pm$11 & -671$\pm$110 & 4766$\pm$226 \\
57845.15         & 4390$\pm$281 & 5450$\pm$1119 & 155$\pm$13 & 77$\pm$18  & -655$\pm$133  & 5403$\pm$374 &     4050$\pm$114 & 5301$\pm$370 & 686$\pm$51   & 383$\pm$31 & -936$\pm$97  & 4371$\pm$118 \\
57935.94$^{(6)}$       & 5008$\pm$862 &   $\cdots$    & 70$\pm$18  & $\cdots$   & 68$\pm$295    & $\cdots$ &    4145$\pm$267 & 6234$\pm$620 & 297$\pm$32   & 295$\pm$32 & -1130$\pm$114 & 4955$\pm$221 \\

\end{tabular}
\end{center}
\textit{Notes:} (2) Full Width Half Maximum; (3) Equivalent Width; (4) shift of the central wavelength with respect to the laboratory wavelength of \ha~and \hb. A negative value corresponds to a blueshift, a positive value to a redshift. (5) In the spectrum taken on 2016 December 10 a cosmic ray hit the CCD at the location of the \hba component, therefore the results of the fit on the \hb~line is not reliable for this epoch. (6) At this epoch, the \hbb component is not present anymore. The last observation (MJD 58\,307.01) was not analysed as the object was back to its "quiescent", pre-outburst state at this epoch.
\end{table}
\end{landscape}


\bsp	
\label{lastpage}
\end{document}